\documentclass[letterpaper,hyper]{JHEP3}
\usepackage[centertags]{amsmath}
\usepackage{array,multirow}
\usepackage{amssymb}    
\usepackage{graphicx}
\usepackage{color}

\usepackage{epsfig}

\def\a{\alpha}

\def\o[#1]{{\rm O}\left({#1}\right)}
\def\dotl[#1,#2]{\left\langle #1, #2 \right\rangle}
\def\dotlb[#1,#2]{[ #1, #2 ]}
\def\dotp[#1,#2]{(#1) \cdot (#2)}
\def\n4sym{{\cal N}=4 SYM}

\def\>{\rangle}
\def\<{\langle}
\def\ads[#1]{$\text{AdS}_{#1}$}
 \usepackage{color}
\newcommand{\norm}[1]{\left| #1 \right|}
\newcommand{\bea}{\begin{eqnarray}}
\newcommand{\eea}{\end{eqnarray}}
\newcommand{\be}{\begin{equation}}
\newcommand{\ee}{\end{equation}}
\newcommand{\ba}{\begin{align}}
\newcommand{\ea}{\end{align}}
\newcommand{\rep}[1]{{\bf{#1}}}
\newcommand{\repb}[1]{{\bf{#1}}}
\newcommand{\repc}[1]{{\boldsymbol{#1}}}
\newcommand{\weyl}[1]{{\boldsymbol{#1}}}
\def\coset[#1,#2]{(\rep{#1}, \rep{#2})}

\title{Correlation Functions in Holographic Minimal Models}
\author{Kyriakos Papadodimas$^a$ and Suvrat Raju$^b$\\
$^a$ Theory Group, Physics Department, CERN, CH-1211 Geneva 23, Switzerland. \\
$^b$ Harish-Chandra Research Institute, Chatnag Marg, Jhunsi, Allahabad 211019, India.}

\date{}

\abstract{We compute exact three and four point functions in the ${\cal W}_N$ minimal models that were recently conjectured to be dual to a higher spin theory in AdS$_3$. The boundary theory has a large number of light operators that are not only invisible in the bulk but grow exponentially with $N$ even at small conformal dimensions. Nevertheless, we provide evidence that this theory can be understood in a ${1 \over N}$ expansion since our correlators look like free-field correlators corrected by a power series in ${1 \over N}$. However, on examining these corrections we find that the four point function of the two bulk scalar fields is corrected at leading order in ${1 \over N}$ through the contribution of one of the additional light operators in an OPE channel. This suggests that, to correctly reproduce even tree-level correlators on the boundary, the bulk theory needs to be modified by the inclusion of additional fields.
 As a technical by-product of our analysis, we describe two separate methods --- including a Coulomb gas type free-field formalism --- that may be used to compute
 correlation functions in this theory.}

\preprint{HRI/ST/1108\\
CERN-PH-TH/2011-}


\begin{document}
\section{Introduction}

A new holographic duality was recently proposed by Gaberdiel and
Gopakumar \cite{Gaberdiel:2010pz}. Their proposal is that certain two
dimensional conformal field theories, the ${\cal W}_N$ minimal models,
are holographically dual to a Vasiliev-type higher spin theory in AdS$_3$ \cite{Vasiliev:2003ev,Vasiliev:1999ba,Bekaert:2005vh,Iazeolla:2008bp,Campoleoni:2009je} coupled to
two complex scalar fields. The boundary CFT can be represented as a
diagonal coset of WZW models of the form
\begin{equation}
\label{coset}
\widehat{su}(N)_k \oplus \widehat{su}(N)_1 \over \widehat{su}(N)_{k+1}
\end{equation}
The semi-classical limit corresponds to taking $k,N\rightarrow \infty$
while keeping the analogue of the 't Hooft coupling
$\lambda\equiv{N\over k+N}$ fixed.

This duality is particularly interesting because the boundary theory
is under good technical control.\footnote{Another interesting proposal for a holographic duality involving solvable large $N$ 2d CFTs is
the one described in \cite{Kiritsis:2010xc}.} The ${\cal W}_N$ CFT is integrable
and correlation functions can, in principle, be computed exactly for
all values of $N$ and $k$. Hence this duality offers a promising
framework where questions in quantum gravity and holography can be
addressed in a quantitative manner. For example, it would be
interesting to study black hole physics, formation and Hawking
evaporation, or even to attempt to ``derive'' the duality by directly
showing how the CFT correlators can be reorganized and formulated as
observables in the dual AdS theory.

For these reasons it is desirable to understand the duality of
\cite{Gaberdiel:2010pz} in more depth. So far, most of the evidence for
it is based on the matching of symmetries and (of part) of the
spectrum of states on both sides. This, as we will discuss, is quite
non-trivial and involves interesting physics \cite{Gaberdiel:2010pz,Gaberdiel:2011wb, Ahn:2011pv,Gaberdiel:2011zw,Gaberdiel:2011nt}. In \cite{Chang:2011mz}, certain 2- and 3-point
functions were computed from both sides and matched, though these
are essentially fixed by the ${\cal W}_N$ symmetry of the theory.
 It would be useful to develop a
systematic formalism that would allow us to compute arbitrary
correlation functions in the ${\cal W}_N$ CFT. In this paper we make
progress in this direction by describing two separate methods that can be used
to do this. The first involves a free-field, Coulomb-gas
representation of the ${\cal W}_N$ CFT. The second reduces computations in the
${\cal W}_N$ CFT to products of computations in ordinary $\widehat{su}(N)$ models. We use these methods to explicitly compute some sample 4-point functions; by doing an OPE expansion of the 4-point functions we are also able to extract
several interesting 3-point functions.

One motivation for computing these correlation functions is a puzzle
in the proposal of \cite{Gaberdiel:2010pz}: the spectrum of the ${\cal
  W}_N$ CFT, in the large $N$ limit, contains at low conformal
dimension the operators that are supposed to be dual to the bulk
fields i.e. the higher spin fields, the two complex scalars and all
their multi-particle combinations. However, in addition to these
states, the CFT contains an ``additional sector'' of states with low
conformal dimension and large entropy. In fact, these states
completely dominate the spectrum in the large $N$ limit since their
entropy (at fixed conformal dimension) grows very fast with $N$. One of the main
goals of this paper is to better understand this ``additional sector'' and its
interactions with the other operators in the theory.

More specifically, we want to understand to what extent the sector
consisting of the higher spin fields and the scalars, which we will
refer to as the ``ordinary sector'' from now on, is decoupled from the
additional sector of states mentioned above. If for example we start with
an initial state where we have turned on a few quanta of the scalar
fields in the bulk, how quickly can they decay into states of the
additional sector? Answering this question is important in order to check
the validity of the duality of \cite{Gaberdiel:2010pz} --- or perhaps to
modify it in some way to account for the additional sector from the bulk
side. One reason that the question is non-trivial is that the number
of light states in the additional sector is extremely large. So even if
the coupling constant between ordinary and additional sector is suppressed
due to large $N$, it is not a priori clear whether the suppression is
strong enough to compensate for the large entropy of additional states.

In \cite{Gaberdiel:2011zw}, it was argued that at $N = \infty$, operators
belonging to the additional sector drop out of the OPE of ordinary sector operators. However, these arguments are insufficient to tell us how fast the
additional sector decouples as $N \rightarrow \infty$ or even whether its couplings to the ordinary
sector are weaker than the self-couplings of the ordinary sector.
The most
direct way to address this
% question of decoupling of the additional sector,
is to actually compute correlation functions at finite $N$ and then
see whether the ordinary and additional sectors decouple as we take
$N\rightarrow \infty$. To do this we mostly focus on the computation
of 4-point functions of the operators dual to the scalar fields in the
bulk. By performing a conformal block decomposition of these
correlators we can see what kind of operators run in the double OPE
and we can isolate the 3-point functions between --- say --- one ``additional'' and two 
``ordinary'' operators. These 3-point functions contain information on
how strongly the additional and ordinary sectors are coupled.

Our computations reveal the following: at $N=\infty$ the ordinary and
additional sectors are indeed decoupled. However, the $N=\infty$ theory is somewhat trivial since all our
evidence suggests that in this limit the theory is free. The couplings
of the ordinary sector to the additional sector vanish at $N = \infty$
but so do the self-couplings of the ordinary-sector.

When we turn on $1/N$
corrections, we find that the two sectors are coupled. Moreover the
coupling between ordinary and additional states starts at the same order in
$1/N$ as the first non-trivial coupling among ordinary states i.e. at
``tree level''. This indicates that the sector of the CFT
corresponding to the higher spin theory in the bulk (together with the
two scalar fields) --- i.e. the ``ordinary sector'' --- is not closed with respect to interactions, {\it even at tree
  level}. Hence {\it it is not possible} to match it to the bulk
higher spin theory, in its original avatar with only two scalars,  beyond the free $N=\infty$ point.

In section \ref{sec:interpret} --- the  last section of our paper --- we start an investigation into whether
the bulk theory can be fixed to reproduce boundary correlation functions, at least at leading order in ${1 \over N}$ (tree level). Our results indicate that this may be possible by adding new degrees of
freedom to the bulk, corresponding to the additional sector of the CFT. We make a tentative proposal for how the lightest
additional-sector state, which we call $\omega$, may be represented as a field with a small negative mass-squared: $m^2_{\omega} \approx -{2 \lambda^2 \over  N}$. This allows a choice of quantization that results in an operator with a very small dimension: $\Delta_{\omega} \approx {\lambda^2 \over N}$  \cite{Klebanov:1999tb}. It would be interesting
to understand how this procedure generalizes to other additional-sector operators and at subleading order in ${\cal O}\left({1 \over N} \right)$.
We leave
this to future work.

Before we close the introduction let us briefly describe how the
computation of the ${\cal W}_N$ correlation functions can be performed
from the technical point of view. As mentioned above, the CFTs under
consideration are integrable. In particular they have a free field
representation \cite{Fateev:1987zh}. The ${\cal W}_N$ CFT can be
realized as a subsector of a theory of $N-1$ free bosons with
background charge. This is the so-called Coulomb-gas or Feigin-Fuchs
representation. The correlation functions of the ${\cal W}_N$ theory
can then be recovered from correlation functions of vertex operators
in a standard free field theory. The only remaining difficulty lies in
the correct insertion of the ``screening operators'' and the evaluation
of certain contour integrals, as we explain in Appendix \ref{sec:coulombgas}.
This method of computing correlation functions was first
applied to the usual $c<1$ minimal models in \cite{Dotsenko:1984nm},
\cite{Dotsenko:1984ad}.

An alternative approach is to reduce the computation of correlators in the coset
to the computation of correlators in ordinary $\widehat{su}(N)$ WZW models using the procedure described in \cite{Douglas:1987cv, Gawedzki:1988hq, Gawedzki:1988nj, Bratchikov:2000mh}. In this procedure, the path integral in the
coset model is written as the path integral in a gauged WZW-model and, by
a change of variables, this can be written as the path integral over three decoupled WZW models.\footnote{As the papers by Gawedzki and Kupiainen \cite{Gawedzki:1988nj,Gawedzki:1988hq} explain, one of these models can be thought of as having
a ``negative level.'' This is not a problem for us
 since its correlators obey the same Knizhnik Zamolodchikov equations as
ordinary WZW models.} Correlation functions in WZW models
can be computed exactly --- at least in the case of 4-point
functions ---  by exploiting the constraints from the
Knizhnik-Zamolodchikov equations in combination with crossing symmetry
\cite{Knizhnik:1984nr}.

Hence we have two different methods to calculate correlation functions
in the ${\cal W}_N$ CFT. We have performed several explicit
computations using both methods and find agreement, as expected of
course.

The plan of the paper is as follows: in section \ref{sec:review} we review the
duality of \cite{Gaberdiel:2010pz} and the puzzle about the additional
states. In section \ref{sec:correlators} we present explicit results for correlators including
the 4-point functions of the scalar
fields in the bulk and study their form in the large $N$ limit. In
section \ref{sec:interpret} we discuss the physical interpretation of our results. The
technical computation of the 4-point functions are presented in the
appendices. In appendix \ref{sec:coulombgas} we calculate the 4-point functions using a
Coulomb-gas formalism. In appendix \ref{sec:bratchikov} we rederive the same results
starting from the correlators of the $\widehat{su}(N)$ WZW models. In
appendix \ref{sec:grouptheory} we summarize some useful group theoretic details.

%%% Local Variables:
%%% mode: latex
%%% TeX-master: "wncorrelators_paper"
%%% End:

\section{Review \label{sec:review}}
\subsection{General Expectations}

One of the most characteristic properties of CFTs with weakly coupled holographic duals is the distinctive form of their spectrum: the number of operators of low conformal dimension is small, in contrast to the large number of operators of high conformal dimension. This property is quite robust and true whether the bulk theory is two-derivative gravity, higher spin gravity or even highly curved (but weakly coupled) string theory (see \cite{ElShowk:2011ag} for a review).
 From the bulk point of view this qualitative form of the spectrum is suggested by the fact that, as expected for any reasonable theory, the number of light fields in AdS stays finite when we take the weak coupling limit $G_N\rightarrow 0$, while --- on the contrary --- the number of black hole microstates blows up, as suggested by the Bekenstein-Hawking entropy formula $S={A\over 4 G_N}$.

The fact that the two types of states (i.e. light fields vs black holes) correspond to different regimes of conformal dimension seems to be quite crucial for the validity of effective field theory (or even perturbative string theory) in the bulk. Given the huge entropy of black hole-type states, one may wonder why they do not dominate the interactions between low energy fields, for example when thinking of them as ``virtual'' intermediate states in scattering processes\footnote{Or more accurately, as intermediate states in the double OPE in the boundary CFT.}. In standard examples of AdS/CFT \cite{Maldacena:1997re,Aharony:1999ti} this does not happen due to the very large mass of the black hole microstates. The exponential suppression due to their mass wins over the enhancement due to the large number of such states.

The theories discussed in \cite{Gaberdiel:2010pz} are the first example of CFTs with a (candidate) holographic dual, where this qualitative form of the spectrum is drastically modified. As we will explain later, at low conformal dimensions the ${\cal W}_N$ CFT contains the small number of operators expected to be dual to the bulk AdS$_3$ fields but in addition it contains an ``additional sector'' of operators with low conformal dimension and very large entropy. The role of these states in the bulk remains quite mysterious.

This makes the duality of \cite{Gaberdiel:2010pz} an interesting case, since it has qualitatively different features from all other known examples of AdS/CFT dualities. What ensures that the presence of an additional sector of light
states with very large entropy does not completely contaminate the
small ``ordinary'' sector of the higher spin theory? The answer lies in the
underlying integrability of the boundary theory. While there are
3-point functions between ordinary and additional states that are nonzero,
these couplings are quite non-generic. For a given choice of ordinary
states only very special additional states can be produced. Most of the
3-point functions are zero, except for very special combinations
allowed by the CFT fusion rules that we review below. Therefore the naive estimate of a
decay rate of ordinary into additional states, where we simply multiply
with the number of final states with the same energy is not
correct. This non-genericity of the couplings is what saves the
ordinary states from being totally mixed with the additional sector.

\subsection{Review of the Spectrum}
Now, we briefly review the spectrum of both sides of the duality proposed
in \cite{Gaberdiel:2010pz}. The duality links a higher spin theory on
\ads[3] to a ${\cal W}_N$ minimal model on the boundary which is defined by a
WZW-coset construction of the form
$$
\widehat{su}(N)_k \oplus \widehat{su}(N)_1 \over \widehat{su}(N)_{k+1}
$$
where we take $k,N \rightarrow \infty$ keeping $\lambda \equiv {N\over k+N}$ fixed. The central charge of the CFT is
\begin{equation}
\label{centralcharge}
\begin{split}
c &=(N-1)\left(1 - {N (N+1) \over p (p+1)} \right)
\end{split}
\end{equation}
where we defined
\begin{equation}
\label{defofp}
p \equiv N + k
\end{equation}

On the bulk side, only the free spectrum of the theory is known. This comprises
two scalars of mass given by
\begin{equation}
  \label{eq:mscalar}
M^2 = -1 + \lambda^2  \ ,
\end{equation}
and a number of other massless gauge fields of spin $2,3, \ldots$ that give
rise to the ${\cal W}_N$ symmetry. The two bulk scalars correspond to two scalar
operators in the CFT of (holomorphic) conformal dimension\footnote{We have $\overline{h}_{\pm} = h_{\pm}$ hence
the full conformal dimension is $\Delta_{\pm} = 1\pm \lambda$.}
\begin{equation}
\label{scalarop}
h_{\pm} = {1\pm \lambda \over 2}
\end{equation}
These values correspond to the two possible quantization of a scalar field of mass \eqref{eq:mscalar}
in AdS$_3$. We denote the two scalars, and the dual operators in the boundary CFT, by $\phi_+$ and $\phi_-$.

The free spectrum of the theory in the bulk consists of multi-particles of the scalar
states and their descendants dressed with the various ${\cal W}_N$ fields. In
fact the partition function of the free bulk theory is given by
\begin{equation}
\label{eq:bulkpartfn}
 Z_{\rm bulk} = (q\overline{q})^{-c/24} Z_{\rm hs} Z_{\rm scal}(h_+)^2 Z_{\rm scal}(h_-)^2
\end{equation}
where
\begin{equation}
Z_{\rm hs} = \prod_{s=2}^{\infty}\prod_{n=s}^\infty \frac{1}{|1-q^n|^2}
\end{equation}
and
\begin{equation}
Z_{\rm scal}(h) = \prod_{j,j'=0}^\infty\frac{1}{1-q^{h+j}\overline{q}^{h+j'}}
\end{equation}
is the usual partition function for a generalized free field of conformal
dimension $h$ on the boundary.

Now, we turn to the boundary. Here, the spectrum of states consists of a number of
``primaries'' of the coset model and each primary is associated with a ``branching function'' that tells us the spectrum of descendants. However, on this side
we know the spectrum {\em exactly} at all values of $N$ and $k$.

Let us describe some qualitative features of the spectrum. Each primary is
labeled by two representations of $SU(N)$, which we denote by $\repb{\Lambda_{+}}$ and $\repb{\Lambda_{-}}$. The allowed representations have the property that the sum of their Dynkin labels\footnote{The Young tableaux corresponding to these Dynkin
labels is made by putting together $d_i$ columns with $i$ boxes.}  is bounded above
\begin{equation}
\sum_{i=1}^{N-1} d_i(\repb{\Lambda_{+}}) < k, \quad \sum_{i=1}^{N-1} d_i(\repb{\Lambda_{-}}) < k + 1,
\end{equation}
The partition function of the boundary theory can be written as
\begin{equation}
  \label{eq:zcftschem}
Z_{\rm CFT}(N,k) = \sum_{\repb{\Lambda_+}, \repb{\Lambda_-}} |b_{(\repb{\Lambda_+};\repb{\Lambda_-})}(q)|^2 \ ,
\end{equation}
where the sum ranges over all the allowed representations indicated above and
the $b$'s are the branching functions. The precise
form of the $b$'s will not be important for us although it is described
in some detail in section 4 of \cite{Gaberdiel:2011zw}.

Qualitatively, what is important is that in the 't Hooft limit, the factor
$(q \overline{q})^{-{c \over 24}} Z_{hs}$ in \eqref{eq:bulkpartfn} comes out automatically as a universal factor in each branching function. We explain this in a little more
technical detail in Appendix \ref{sec:grouptheory}. So, what is of interest
is mainly the spectrum of primaries.

The dimension of a primary is given by
\begin{equation}
\label{eq:dimprim}
\Delta\coset[\Lambda_{+},\Lambda_{-}] = {\left[(p + 1) (\repb{\Lambda_{+}} + \weyl{\rho}) - p (\repb{\Lambda_{-}} + \weyl{\rho})\right]^2 - \weyl{\rho}^2  \over  p (p + 1) }
\end{equation}
where $\weyl{\rho}$ is the Weyl vector of $SU(N)$ and the ``square'' of a weight vector is taken using the inner product described in Appendix \ref{sec:grouptheory}.

The two scalars $\phi_+$ and $\phi_-$ in this notation correspond to
\begin{equation}
\phi_+ \equiv \coset[f,1], \quad \phi_- \equiv \coset[1,f]
\end{equation}
where $\rep{f}$ denotes the fundamental representation of $SU(N)$ and $\rep{1}$ the trivial representation. For their conjugates we use the obvious notation $\overline{\phi}_+ \equiv (\overline{\rep{f}},\rep{1})\,,\,\overline{\phi}_-\equiv (\rep{1},\overline{\rep{f}})$,
where $\rep{\overline{f}}$  indicates the anti-fundamental representation.
The primaries corresponding to the scalars $\phi_+$ and $\phi_-$ have dimensions
\begin{equation}
\begin{split}
\label{dimensions}
\Delta_+ &= {((p+1) \repb{f} + \weyl{\rho})^2 - \weyl{\rho}^2 \over  p (p+1)} = {(N-1) (N+1+p) \over  N p} \approx 1 +\lambda \\
\Delta_- &= {(p \rep{f} - \weyl{\rho})^2 - \weyl{\rho}^2 \over  p (p+1)} = {(N-1) (p - N) \over  N (p + 1)} \approx 1 -\lambda \\
\end{split}
\end{equation}
In \cite{Gaberdiel:2011zw}, it was shown that it is possible to find all the states in
\eqref{eq:bulkpartfn} in the CFT partition function \eqref{eq:zcftschem}.

However, as we point out now, the CFT also has additional states that are not naively visible in
the bulk. Let us focus for simplicity on the large class of operators for which $\repb{\Lambda_{+}} = \repb{\Lambda_{-}}.$ For these operators, we have
\begin{equation}
\label{eq:dimsimple}
\Delta \coset[\Lambda_{+},\Lambda_{-}] = {2 C_2 (\rep{\Lambda_{+}}) \over p (p+1)},
\end{equation}
where $C_2$ is the ``second Casimir.''

Now we immediately see that we have a puzzle. The spectrum of states in
the partition function \eqref{eq:bulkpartfn} consists of operators
with dimensions
\[
n_{+} h_{+} + n_{-} h_{-} + m
\]
where $n_{\pm}$ and $m$ are integers.
On the other hand \eqref{eq:dimsimple} can result in all sorts of fractional
dimensions. Moreover, for operators where the second Casimir is small compared
to $p (p + 1)$ --- this is a large subclass --- equation \eqref{eq:dimsimple} gives dimensions that are close to $0$,
in the large $N$ limit.

This implies that the CFT has a very large class of light states that do not
seem to be present in the bulk theory. It is also possible to discern
the following facts about these states\footnote{This was worked out in collaboration with Shiraz Minwalla and Mukund Rangamani.}
\begin{enumerate}
\item
The light states do not form ``small bands'' about the discrete states
corresponding to the scalar Fock space. In fact, in the subclass of states
with $\repb{\Lambda_{+}} = \repb{\Lambda_{-}}$ we already have primaries with dimensions as large as ${k N \over 4(k + N)}$.
\item
The light states are very dense. In fact, even if we restrict just to
the subector above, the number of primaries below some
fixed conformal dimension $\Delta$ (much smaller than $N$) scales like $\exp{\left[{\pi \over \lambda} \sqrt{4 N \Delta \over 3}\right]}$!
\end{enumerate}

Hence, we see that the boundary CFT naively violates the expectations
from a CFT with a bulk dual. This leads us to the question of
decoupling.

\subsection{Decoupling}
Naively the presence of such a large number of additional states would
lead us to suspect that it is impossible to write down a theory that just
describes the dynamics of $\phi_+$ and $\phi_-$. This difficulty is somewhat
ameliorated by the following fact.

In the 't Hooft limit, if we consider the OPE of two primaries of the CFT with representations
$(\repb{\Lambda_{+}}^1, \repb{\Lambda_{-}}^1)$ and $(\repb{\Lambda_{+}}^2, \repb{\Lambda_{-}}^2)$ where all
the Dynkin indices are much smaller than $k$ then
only a very restricted set of operators can appear on the right hand side:
\begin{equation}
\begin{split}
  \label{eq:fusion}
  \phi_{(\repb{\Lambda_{+}^1},\repb{\Lambda_{-}^1})}(x,\overline{x}) &\,\cdot\, \phi_{(\repb{\Lambda_{+}^2},\repb{\Lambda_{-}^2})}(0)\\ &= \sum_{\tiny \begin{array}{l} \repb{\Lambda_{+}^3} \in \repb{\Lambda_{+}^1} \otimes \repb{\Lambda_{+}^2}, \\ \repb{\Lambda_{-}^3} \in \repb{\Lambda_{-}^1} \otimes \repb{\Lambda_{-}^2} \end{array}} {1 \over \norm{x}^{\Delta_1 + \Delta_2 - \Delta_3}}  C^{(\repb{\Lambda_{+}^3},\repb{\Lambda_{-}^3})}_{(\repb{\Lambda_{+}^1},\repb{\Lambda_{-}^1}), (\repb{\Lambda_{+}^2},\repb{\Lambda_{-}^2})}(0)\left( \phi_{(\repb{\Lambda_{+}^3},\repb{\Lambda_{-}^3})}(0) + \ldots \right),
  \end{split}
\end{equation}
where $\Delta_m$ denotes the dimension of the representation $(\repb{\Lambda_+}^m, \repb{\Lambda_{-}}^m)$, the $\ldots$ denotes the coset descendants and the tensor product is
the ordinary $SU(N)$ tensor product.

For example, in the OPE of $\phi_+$ with its conjugate we can get exactly
two representations
\begin{equation}
\phi_+(x,\overline{x}) \,\cdot\,\overline{\phi}_+(0) = C^{\coset[1,1]}_{\phi_+ \overline{\phi}_+} \left(\phi_{\coset[1,1]}(0) + \ldots \right) +  C^{\coset[adj,1]}_{\phi_+ \overline{\phi}_+} \left(\phi_{\coset[adj,1]}(0) + \ldots \right),
\end{equation}
where $\rep{adj}$ denotes the adjoint representation of $SU(N)$. From \eqref{eq:dimprim} we see that the operator $\phi_{\coset[adj,1]}(0)$ has  dimension
\begin{equation}
  \label{eq:dimadj}
  \Delta_{\coset[adj,1]} = 2 +  {2N \over p} = 2\left(1 + \lambda\right) = 2\Delta_{({\rep{f},\rep{0}})}.
\end{equation}
This dimension formula is highly suggestive and tells us that we should
identify  $\phi_{\coset[adj,1]}(0)$ as a ``double trace'' operator\footnote{We define the normal-ordered product
of two operators as the non-singular part of the OPE in the limit where the two points are brought together. Obviously this only makes sense for free fields, or in theories with a large $N$ expansion (i.e. for generalized free fields).
In this paper, we will often put the normal ordering symbol around operators but this is mainly meant as an intuitive guide. Whenever we compute a correlation function, we define the operator precisely using the coset construction.}
\begin{equation}
\begin{split}
\phi_{\coset[adj,1]}(x) \sim\, &\colon\phi_+ \overline{\phi}_+\colon(x) \\
\end{split}
\end{equation}

In fact, working through \eqref{eq:fusion} we find that  we can consider higher point functions
of $\phi_+(x)$ and its conjugate and as long as we keep the number
of operators small compared to $N$ and $k$, we can identify the representations
that occur in the fusion products as ``multi-trace'' products of $\phi_+$
and its conformal descendants. A very similar story holds for $\phi_-$. This implies that the ``additional'' light states do not appear at all in the computation of these correlation functions. Hence, at the level of correlators on the plane we can consider the dynamics of $\phi_+$  completely independently of $\phi_-$ and the other light states. This was the basis of the modified
duality proposed in \cite{Chang:2011mz}.

However, the theory comprising only $\phi_+$ cannot be consistent as a
complete CFT. For one, if we consider correlators of ${\cal O}(k)~\phi_+$ operators, we will discover that there are other states in the theory beyond $\phi_+$ and its multi-traces. Second, this sector is not modular invariant.

The original duality proposed in \cite{Gaberdiel:2010pz} consisted of both
scalars $\phi_+$ and $\phi_-$. The OPE
of these two fields reads
\begin{equation}
\label{eq:phi12fusion}
\phi_{\coset[f,1]}(x,\overline{x}) \,\cdot\,\phi_{\coset[1,f]}(0) =
C_{\coset[f,1],\coset[1,f]}^{\coset[f,f]} \left(\phi_{\coset[f,f]}(0) + \ldots \right).
\end{equation}
In this paper we will use
\begin{equation}
\label{defomega}
\omega \equiv \phi_{\coset[f,f]}
\end{equation}
following the notation of \cite{Gaberdiel:2010pz}.

When we consider the dimension of $\omega$, we find a puzzle:
\begin{equation}
  \label{eq:dimomega}
  \Delta_{\omega} = {N^2 - 1 \over N p (p + 1)} \approx {\lambda^2 \over N},
\end{equation}
which vanishes in the 't Hooft limit! However this is not the only puzzling
feature. Where is the double trace operator $\colon\phi_+ \phi_-\colon$ in the OPE \eqref{eq:phi12fusion}, that we would expect in a weakly coupled theory?

It turns out that the double trace operator is a {\em conformal descendant}
of $\omega$. Defining
\begin{equation}
\label{defpsi}
\psi \equiv {1 \over \Delta_\omega} [L_{-1},[ \overline{L}_{-1}, \omega]] = {1\over \Delta_\omega} \partial \overline{\partial}\omega,
\end{equation}
we have
\begin{equation}
\psi(x) \sim \colon \phi_+ \phi_- \colon(x),
\end{equation}

Moreover, at {\em infinite $N$} the representation
$\coset[f,f]$ becomes reducible and we find that $\omega$ drops out of the fusion
product \eqref{eq:phi12fusion} leaving behind only the double trace operator.
So, exactly at $N = \infty$, it is possible to consider correlators of $\phi_+$, $\phi_-$ and their multi-traces without worrying about the light states.

However, the {\em infinite $N$} theory is somewhat boring because it is a free theory. All correlators are disconnected and can be written as products of
two point functions. For example, at infinte $N$,
\begin{equation}
\langle \phi_+(x) \overline{\phi}_+(0) \phi_{\coset[adj,1]}(y) \rangle = \langle \phi_+(x) \overline{\phi}_+(y) \rangle  \langle \phi_+(y) \overline{\phi}_+(0) \rangle.
\end{equation}

In this paper, we would like to study the finite but large $N$ theory.
The question we would like to address is to what extent can
we consider correlation functions of $\phi_+$ and $\phi_-$ without worrying
about the extra light states.

We will address this by computing three
point functions. We will find below that at finite but large $N$, correlation functions have an expansion in terms of  ${1 \over \sqrt{N}}$: this means that
three point functions scale like ${1 \over \sqrt{N}}$, and the {\em connected} part of a 4-point function scales like  ${1 \over  N}$ and so on.

However, we find that the following fields from the ``additional'' sector\footnote{We have chosen the normalization
of the operators $\psi_L, \psi_R$ so that their 2-point function is order 1.}
\begin{equation}
\label{firstdesc}
\psi_L \equiv {1\over \sqrt{\Delta_\omega}} [L_{-1}, \omega],\qquad
\psi_R \equiv {1\over \sqrt{\Delta_\omega}} [\overline{L}_{-1}, \omega]
\end{equation}
couple to $\phi_+$ and $\phi_-$ at
{\em leading order} i.e
\begin{equation}
C_{\phi_+ \phi_-}^{\psi_L} = C_{\phi_+ \phi_-}^{\psi_R} \approx {1 \over \sqrt{N}}
\end{equation}
We will first show how this result is derived in section \ref{sec:correlators} and then discuss
its implications in section \ref{sec:interpret}.

%%% Local Variables:
%%% mode: latex
%%% TeX-master: "wncorrelators_paper"
%%% End:

\section{Explicit Results for Correlators}
\label{sec:correlators}

We are interested in computing four point functions of the bulk
scalars using the boundary theory. These scalars are dual to operators
that we denote by the same symbols --- $\phi_{+}$ and $\phi_{-}$. In computing a 4-point function we can use conformal invariance to
place 3 points at $(0,1,\infty)$ and then the answer depends
only on the position $x$ of the fourth point\footnote{For notational simplicity sometimes we only write ``$x$'' as the
argument of the correlator, but in most cases the correlators depend both on $x$ and $\overline{x}.$} (or equivalently we can
think of $x$ as a conformal cross ratio). In this section, we will first
present answers for  the following three correlators
\begin{align}
\label{firstcorrelator} &G_{\phi_{+}\phi_{+}}(x) \equiv \langle \phi_{+}(\infty) \overline{\phi}_{+}(1) \phi_{+}(x)
\overline{\phi}_{+}(0)   \rangle, \\
\label{secondcorrelator} &G_{\phi_{-}\phi_{-}}(x) \equiv \langle \phi_{-}(\infty) \overline{\phi}_{-}(1) \phi_{-}(x)
\overline{\phi}_{-}(0)   \rangle, \\
\label{thirdcorrelator} &G_{\phi_{+} \phi_{-}}(x) \equiv \langle \phi_{-}(\infty) \phi_{+}(1) \overline{\phi}_{+}(x)
\overline{\phi}_{-}(0) \rangle.
\end{align}
Each of these correlators is of specific interest. The first two correlators
are useful in telling us that $\phi_{+}$ and $\phi_{-}$ are generalized free-fields\footnote{i.e. that their
correlators factorize at large $N$.}. The third correlator is of the most interest since by doing an OPE
expansion, we can extract the three-point couplings of $\phi_{+}$ and $\phi_{-}$
to the lightest additional-sector field --- $\omega$.

We then present the answers for two correlators that involve not just the
basic fields $\phi_{\pm}$ but also ``multi-trace'' operators of these fields. In the language of the coset, these involve higher representations.
\begin{equation}
\label{correlators45}
\begin{split}
G_{\phi_{-}^2\phi_{+} }(x) \equiv  \langle \phi_{-}^2(\infty) \phi_{+}(1) \overline{\phi}_{+}(x) \overline{\phi}_{-}^2(0) \rangle. \\
G_{ \phi_{+}^2\phi_{-}}(x) \equiv  \langle \phi_{+}^2(\infty) \phi_{-}(1) \overline{\phi}_{-}(x) \overline{\phi}_{+}^2(0) \rangle. \\
\end{split}
\end{equation}
where by $\phi_{-}^2$ and $\phi_{+}^2$ we mean the following coset primary fields
\begin{equation}
\phi_{-}^2 \equiv \coset[1,asym], \quad \phi_{+}^2 \equiv \coset[asym,1]
\end{equation}
These correlators are of interest because they show that not only $\omega$ but even other operators in the
additional sector, such as $\omega'_{\pm}$ defined in \eqref{defomegaprime},
also couple to ordinary sector operators.

In the appendices, we outline two separate procedures to compute these
correlators. In Appendix A we set up a Coulomb gas like
formalism for the coset.  In Appendix B we use another
prescription given by Gawedzki and Kupiainen \cite{Gawedzki:1988hq}
and elaborated by Bratchikov \cite{Bratchikov:2000mh}. This reduces
the computation of coset correlators to the computation of correlators
in the product WZW model $\widehat{su}(N)_k \oplus \widehat{su}(N)_1$ and
another correlator in a ``ghost'' WZW model which has level
$k^* = -k-2N-1$. All these correlators can be obtained by solving the
Knizhnik-Zamolodchikov equation. As we show in the appendices, the answers obtained by
the two prescriptions agree precisely.

We also perform an important physical check on the three point
functions that we obtain below. Since we have identified generalized
free fields on the boundary $\phi_{+}$ and $\phi_{-}$ and the
corresponding ``double trace'' operator $\colon\phi_{+} \phi_{-}\colon$, it is
natural to expect that at leading order in the ${1 \over \sqrt{N}}$
expansion, the double trace operator appears with coefficient $1$ in
the OPE of $\phi_{+}$ and $\phi_{-}$. This combined with the fusion
rules described above, in principle, fixes the OPE coefficients of all
operators in this coset representation to leading order in ${1 \over
  \sqrt{N}}$. Using this we check that the field $\omega$, which is a
conformal ``ancestor'' of the double trace operator has the three
point function that we compute using the four point correlators above.

Let us briefly explain how we choose the normalization of operators. We choose
the normalization of a "single-trace" conformal primary ${\cal O}$ of dimension
$\Delta$, in such way that
$$
\langle {\cal O}(x) \overline{\cal O}(0)\rangle = {1\over |x|^{2\Delta}}
$$
In particular, this will be the normalization of the single-trace operators $\phi_+$ and $\phi_-$.
For double-trace conformal primary operators the normalization is fixed by defining the double-trace operator
as the non-singular part of the OPE of two single-trace operators when they are brought to the same point\footnote{\label{normdouble} For
example this implies that (at large $N$) we have
$
\langle \colon {\cal O}{\cal O}\colon(x)\, \colon\overline{\cal O}\overline{\cal O}\colon(0) \rangle = {2\over |x|^{4\Delta}}
$
but on the other hand, for ${\it different}$ single trace operators ${\cal O}_1,{\cal O}_2$ we have
$
\langle \colon{\cal O}_1 {\cal O}_2\colon(x)\, \colon\overline{\cal O}_1\overline{\cal O}_2\colon(0) \rangle = {1\over |x|^{2(\Delta_1+\Delta_2)}}
$.}.

\subsection{Answers for four point functions \label{list4pt}}

\paragraph{Four point function of $\phi_{+}$:\\}
In Appendix A, it is shown that this four point function is given by:
\begin{equation}
\label{firstcorr}
\begin{split}
G_{\phi_+ \phi_+}(x) = \left| x (1 - x) \right|^{-2 \Delta_+} \Bigg[
  &\left|(1-x)^{k+2N \over k+N}\,
  _2F_1\big(\frac{k+N+1}{k+N},-\frac{1}{k+N};-\frac{N}{k+N};x\big)\right|^2
  \\ &+ {\cal N}_1 \left|x^{k+2N \over k+N}\,
  _2F_1\big(\frac{k+N+1}{k+N},-\frac{1}{k+N};\frac{2k + 3
    N}{k+N};x\big)\right|^2 \Bigg],
\end{split}
\end{equation}
with
\begin{equation}
\begin{split}
{\cal N}_1 &= -{\Gamma({k + 2 N - 1 \over k + N}) \Gamma
  \big(\frac{-N}{k+N}\big)^2 \Gamma \big(\frac{2 k+3 N+1}{k+N}\big)
  \over \Gamma({-k - 2N - 1 \over k+N}) \Gamma
  \big(\frac{1-N}{k+N}\big) \Gamma \big(\frac{2 k+3 N}{k+N}\big)^2
  \big)}
\end{split}
\end{equation}
where we remind the reader that the conformal dimension $\Delta_+$ is given by \eqref{dimensions}.

\paragraph{Four point function of $\phi_{-}$: \\}
In the Appendix, it is shown that this four point function is given by:
\begin{equation}
\label{secondcorr}
\begin{split}
G_{\phi_- \phi_-}(x) =  \left|(1-x)  x \right|^{-2 \Delta_{-}} \Bigg[  &\left| (1 - x)^{1+k \over k+N+1} \,
   _2F_1\big(\frac{k+N}{k+N+1},\frac{1}{k+N+1};\frac{N}{k+N+1};x\big)\right|^2 \\ &+ {\cal N}_2  \left| x^{\frac{1+k}{k+N+1}} \, _2F_1\big(\frac{k+2}{k+n+1},\frac{2
   k+N+1}{k+N+1}; \frac{2
   k+N+2}{k+N+1};x \big) \right|^2  \Bigg]
\end{split}
\end{equation}
with $\Delta_-$ given by \eqref{dimensions}.
\begin{equation}
\begin{split}
{\cal N}_2
&= -{ \Gamma
   \big(\frac{N}{k+N+1}\big)^2 \Gamma \big(\frac{2
   k+N+1}{k+N+1}\big)  \Gamma({k + 2 \over k + N + 1}) \over \Gamma
   \big(\frac{N-1}{k+N+1}\big) \Gamma \big(\frac{2 k+N+2}{k+N+1}\big)^2 \Gamma({-k \over k + N + 1})}
\end{split}
\end{equation}

\paragraph{Mixed four point function of $\phi_{+}$ and $\phi_{-}$:\\}
We now turn to the most interesting case. In the appendix, it is shown
that the mixed correlator is given by a remarkably simple expression
\begin{equation}
\label{mixed4pt}
\begin{split}
G_{\phi_{+}\phi_{-}}(x) &=
|1 - x|^{-2 \Delta_+} |x|^{2\over N} \left|1 + {1 - x \over N x} \right|^2
\end{split}
\end{equation}

\paragraph{Double trace of $\phi_{-}$ with $\phi_{+}$: \\}
This answer is also given by a very simple expression:
\begin{equation}
G_{ \phi_{-}^2\phi_{+}}(x) =  2 \left|1-x\right|^{-2\Delta_+} |x|^{\frac{4}{N}}
\left|1 + 2{1-x \over N x}\right|^2,
\end{equation}

\paragraph{Double trace of $\phi_{+}$ with $\phi_{-}$:\\}
This answer is almost identical to the expression above:
\begin{equation}
G_{ \phi_{-}^2\phi_{+}}(x) =  2 \left|1-x\right|^{-2\Delta_-} |x|^{\frac{4}{N}}
\left|1 + 2{1-x \over N x}\right|^2,
\end{equation}

\subsection{Limiting behaviour of correlators at large $N$ and small distance}
To gain some intuition for these answers, let us expand these
answer in various limits. In the large $N$ limit, we expect that the
fields $\phi_{+}$ and $\phi_{-}$  should become free. So, all the
correlators computed above should break up into a sum of a product of
two point functions. Starting with the four point function of $\phi_+$ we
find
\begin{equation}
G_{\phi_{+}\phi_{+}}(x) \underset{N \rightarrow
  \infty}{\longrightarrow} \left(|x|^{-2(\lambda+1)} +
|1-x|^{-2(\lambda +1)} \right)+ {\cal O}(1/N)
\end{equation}
which is precisely what we expect for a correlator of the form
$\langle \phi \overline{\phi} \phi \overline{\phi}\rangle$ for a
generalized free field $\phi$ of dimension $\lambda+1$.

We can also check that if we take $x \rightarrow 0$ (even at finite $N$), the correlator is
dominated by the OPE channel where the identity runs in between the
two operators on the left and the two on the right:
\begin{equation}
\label{firstsmallx}
G_{\phi_{+}\phi_{+}} \underset{x\rightarrow 0}{\longrightarrow}
\norm{x}^{-2 \Delta_+}.
\end{equation}
To see this we merely need to note that,
in this limit, the correlator is dominated by the term containing the
first hypergeometric function in \eqref{firstcorr}. The hypergeometric
function itself becomes $1$ at $x = 0$, leaving behind the
pre-factors, which yield the behaviour \eqref{firstsmallx}. In fact,
one can check that a similar expansion exists near $x \rightarrow 1$,
although it is not manifest in the form \eqref{firstcorr}.

Turning now to the four point function of $\phi_{-}$, we find that in
 the large $N$ limit
\begin{equation}
G_{\phi_{-}\phi_{-}}(x) \underset{N \rightarrow \infty}{\longrightarrow} \left(|x|^{-2(1-\lambda)} + |1-x|^{-2(1-\lambda)}  \right)+ {\cal O}(1/N)
\end{equation}
which is the 4-point function of a generalized free field of $\Delta_{-} = 1-\lambda$, as expected.
We can also check, as above,  that if we take $x \rightarrow 0$, the correlator is dominated by the OPE channel where the identity runs in between the two operators on the left and the two on the right:
\begin{equation}
\label{secondsmallx}
G_{\phi_{-}\phi_{-}} \underset{x\rightarrow 0}{\longrightarrow} \norm{x}^{-2 \Delta_{-}}.
\end{equation}

Turning now to the mixed four point function of $\phi_{+}$ and $\phi_{-}$,
we see that this correlator manifestly has the correct large $N$ behaviour. In the large
$N$ limit, we have
\begin{equation}
G_{\phi_+ \phi_-}  \underset{N \rightarrow \infty}{\longrightarrow} |1-x|^{-2(1 + \lambda)},
\end{equation}
which is exactly what we expect. Moreover, this tells us that in the large $N$
limit $\phi_{+}$ and $\phi_{-}$ are weakly coupled.
On the other hand, as we take $x \rightarrow 1$, we pick up the most singular
term in $1-x$, which again leads to
\begin{equation}
G_{\phi_+ \phi_-}  \underset{x \rightarrow 1}{\longrightarrow} \norm{1 - x}^{-2 \Delta_{+}}.
\end{equation}

The correlators of what we have called $\phi_{-}^2$ with $\phi_{+}$ are also
of interest.  We can check that
\begin{equation}
G_{\phi_{-}^2 \phi_{+}} \underset{N \rightarrow \infty}{\longrightarrow} 2|1-x|^{-2 (1 + \lambda)}
\end{equation}
which is exactly what we expect. Notice that the factor of 2 is coming from the 2-point function of the
double trace operator $\phi_{-}^2$, as mentioned in footnote \eqref{normdouble}.
Second, we can also check that
\begin{equation}
G_{\phi_{-}^2 \phi_{+}} \underset{x \rightarrow 1}{\longrightarrow}2
|1-x|^{-2 \Delta_+},
\end{equation}
which is a sign of the fact that as we take $x \rightarrow 1$ the $\phi_{+}$ and $\overline{\phi}_{+}$ inside the
correlator fuse to give the identity with coefficient 1.

The results for the correlator of $\phi_{+}^2$ with $\phi_{-}$ are very similar. We have:
\begin{align}
&G_{\phi_{+}^2 \phi_{-}} \underset{N \rightarrow \infty}{\longrightarrow}2 |1-x|^{-2 (1 - \lambda)} \\
&G_{\phi_{+}^2 \phi_{-}} \underset{x \rightarrow 1}{\longrightarrow}2
|1-x|^{-2 \Delta_-},
\end{align}

\subsection{The three point function $C^{\omega}_{\phi_{+} \phi_{-}}$}
With these four point functions in hand, we can now extract the
three point functions in the coset model. We focus on the most
interesting three point function, which is $C^{\omega}_{\phi_{+} \phi_{-}}$ --- since this tells us the coupling between the fundamental scalars in the
theory and the prototypical hidden sector field, $\omega$.

From \eqref{eq:fusion}, as we take $x \rightarrow 0$ in \eqref{mixed4pt}, we need to consider the fusion of $\phi_{+}$ and $\phi_{-}$ and only {\em one} coset representation appears on the right hand side. We remind the reader
that the primary field of this representation is denoted by
\begin{equation}
\omega \equiv (\rep{f},\rep{f}).
\end{equation}
with a dimension $\Delta_{\omega} = {N^2 - 1 \over N p (p +1)} \approx {\lambda^2 \over N}$
The other operators that will be important to us are the following conformal descendants of the primary $\omega$
\begin{equation}
\label{defpsilr}
 \psi_L \equiv {1\over \sqrt{\Delta_\omega}} [L_{-1}, \omega]\qquad ,\qquad
\psi_R \equiv {1\over \sqrt{\Delta_\omega}} [\overline{L}_{-1}, \omega]
\end{equation}
and $\psi$ which was already defined in \eqref{defpsi}
\begin{equation}
\label{boundaryconstraint}
\psi \equiv {1 \over \Delta_\omega} [L_{-1},[ \overline{L}_{-1}, \omega]]
\end{equation}
We have chosen these normalizations so that all three operators $\psi_L, \psi_R, \psi$ are normalized to have a unit two-point function.
As we mentioned above the operators $\omega,\psi_L,\psi_R$ belong to the additional sector while
$\psi$ to the ordinary sector. Moreover $\psi$ has an important
physical interpretation as the ``double trace'' operator of
$\phi_{+}$ and $\phi_{-}$ as we explore below.

Now, we take the point $x$ close to $0$. In that case, we can do an OPE expansion
\begin{equation}
\begin{split}
\overline{\phi}_{+}(x,\overline{x})\,\cdot\, \overline{\phi}_{-}(0,0) = |x|^{{2 \over N}-2}\Big( &
\overline{C}_{\phi_{+} \phi_{-}}^\omega  \,\overline{\omega}(0)  + x\, \overline{C}_{\phi_{+}
\phi_{-}}^{\psi_L} \,\overline{\psi}_L(0) \cr & + \overline{x}\, \overline{C} _{\phi_{+} \phi_{-}}^{\psi_R}\,
\overline{\psi}_R(0)  + |x|^2\,\overline{C}_{\phi_{+} \phi_{-}}^{\psi} \, \,\overline{\psi}(0)
+\ldots\Big)
\end{split}
\end{equation}
If we substitute this back into \eqref{mixed4pt}, we are left with several three point functions. The three-point
functions that we need are
\be
\begin{split}
&\langle \overline{\omega}(0) \phi_{+}(1) \phi_{-}(\infty) \rangle = C^{\omega}_{\phi_{+} \phi_{-}}, \\
&\langle \overline{\psi}_L(0) \phi_{+}(1) \phi_{-}(\infty) \rangle = \langle \overline{\psi}_R(0) \phi_{+}(1) \phi_{-}(\infty) \rangle  = { \left(\Delta_{\omega} + \Delta_{+} - \Delta_{-} \right) \over 2 \sqrt{\Delta_{\omega}}} C^{\omega}_{\phi_{+} \phi_{-}}, \\
&\langle \overline{\psi}(0) \phi_{+}(1) \phi_{-}(\infty) \rangle = {\left(\Delta_{\omega} + \Delta_{+} - \Delta_{-} \right)^2  \over 4 \Delta_{\omega}} C^{\omega}_{\phi_{+} \phi_{-}} \\
\end{split}
\ee

So, we expect that the four point
function will have the behaviour
\begin{equation}
\begin{split}
\langle \phi_{-}(\infty) \phi_{+}(1)\overline{\phi}_{+}(x) \overline{\phi}_{-}(0)\rangle
=  C_{\phi_{+} \phi_{-}}^\omega |x|^{{2 \over N}-2}\Big(& C_{\phi_{+} \phi_{-}}^\omega + { \Delta_{\omega} + \Delta_{+} - \Delta_{-}   \over 2 \sqrt{\Delta_{\omega}}}  \left(x\,
C_{\phi_{+} \phi_{-}}^{\psi_L}  + \overline{x}\,
C_{\phi_{+} \phi_{-}}^{\psi_R} \right) \\ &+  { \left(\Delta_{\omega} + \Delta_{+} - \Delta_{-} \right)^2 \over 4 \Delta_{\omega}} |x|^2 \, C_{\phi_{+} \phi_{-}}^{\psi}+\ldots\Big)
\end{split}
\end{equation}

As we take $x \rightarrow 0$, the four point function \eqref{mixed4pt} does have
this form. This allows us to read off the exact OPE coefficients. Defining the auxiliary quantities
\be
c^{\omega} = {1 \over \sqrt{N}}, \quad c^{\psi} = {1 \over N} \sqrt{(N^2 - 1) (p + 1) \over p},
\ee
we find that the OPE coefficients are given by
\be
\begin{split}
& C_{\phi_{+} \phi_{-}}^\omega= (c^{\omega})^2,\\
& C_{\phi_{+} \phi_{-}}^{\psi_L} =  C_{\phi_{+} \phi_{-}}^{\psi_R}  = c^{\omega} c^{\psi}, \\
& C_{\phi_{+} \phi_{-}}^{\psi} = (c^{\psi})^2.
\end{split}
\ee
It is the 't Hooft limit that is of relevance for the bulk, and in this limit we find
\begin{equation}
\label{opecoeffs}
\begin{split}
& C_{\phi_{+} \phi_{-}}^\omega= {1 \over N}. \\
& C_{\phi_{+} \phi_{-}}^{\psi_L} = C_{\phi_{+} \phi_{-}}^{\psi_R} \approx {1 \over \sqrt{N}}, \\
& C_{\phi_{+} \phi_{-}}^\psi \approx 1, \\
\end{split}
\end{equation}
Although the reader might find it a little counter-intuitive that different operators in the same conformal family
contribute with different powers of $N$, this is entirely consistent with conformal invariance as we show below in subsection \eqref{computationconformal}. This phenomenon is a consequence of the fact that the dimension of $\omega$ vanishes in the large $N$ limit. In fact, the OPE coefficients \eqref{opecoeffs} are
in fact, very natural even from a physical viewpoint as we now discuss.

\subsubsection{$C^{\omega}_{\phi_{+} \phi_{-}}$ from conformal invariance \label{computationconformal}}
This question we ask in this subsection is the following. Let us say we
knew nothing about how to compute four point correlation functions. Could
we still guess at least the leading order three-point coefficients $\eqref{opecoeffs}$? Surprisingly, we will show that the answer is ``yes''. We show below that the results
\eqref{opecoeffs}  follow immediately from a simple physical assumption and conformal invariance. We have written this subsection to be logically independent
of the other computations in this section, and consequently it involves
some repetition.

The physical assumption is that the operator  $\psi$, defined in \eqref{defpsi} is a ``double trace''
operator:
\begin{equation}
\label{psidouble}
\psi(x) \sim \colon \phi_{+}\phi_{-}\colon(x).
\end{equation}
Note that $\psi$ has precisely the right conformal dimension for this,
since
\begin{equation}
\Delta_{\psi} = \Delta_{+} + \Delta_{-} - {2 \over N}.
\end{equation}
This physical assumption tells us that the operator $\psi$ should appear
with coefficient $1$ to leading order in ${1 \over N}$ in the OPE of $\phi_{+}$ and $\phi_{-}$:
\begin{equation}
\phi_{+}(x,\overline{x}) \,\cdot \,\phi_{-}(0) = \psi(0) + \ldots+ {\cal O}\left({1 \over N}\right)
\end{equation}

With this physical assumption, conformal invariance fixes the other OPE coefficients listed in \eqref{opecoeffs} to leading order in ${1 \over N}$. This is  because $\psi$ is a conformal descendant of $\omega$, and so its OPE coefficient is fixed in terms of the OPE coefficient of $\omega$.

On general grounds the OPE has the form
\begin{equation}
\label{ope2}
\phi_+(x) \phi_-(0) = {C_{\phi_+ \phi_-}^\omega   \over |x|^{\Delta_{+} + \Delta_{-} - \Delta_{\omega}}}
\left(\omega(0) + x\, A \,(\partial \omega)(0) + \overline{x} \,A \,(\overline{\partial}\omega)(0) + |x|^2 \,B\,\partial \overline{\partial}\omega(0)+\ldots  \right)
\end{equation}
where the coefficients $A,B,...$ are completely determined by kinematics of the conformal group. We quickly review
how exactly this can be done.

The easiest way to fix these coefficients is to demand consistency of the OPE with
the exact form of the 3-point functions of the operators
\begin{equation}
\label{threept12omega}
\langle \phi_+(x) \phi_-(0) \overline{\omega}(y) \rangle = C_{\phi_+ \phi_- \overline{\omega}} {1 \over |x|^{\Delta_{+} + \Delta_{-} -
\Delta_{\omega}}} {1 \over |y|^{\Delta_{-} +  \Delta_{\omega} - \Delta_{+}}}
{1 \over |y-x|^{\Delta_{+} + \Delta_{\omega} - \Delta_{-}}}
\end{equation}
We take $y > x$, so that we can continue to use \eqref{ope2}. Notice that we are working in a normalization
of $\omega$ where its 2-point function is
\begin{equation}
\label{twopoint}
\langle \omega(x) \overline{\omega}(0)\rangle = {1\over |x|^{2\Delta_\omega}}
\end{equation}
which implies that the OPE coefficients and 3-point functions are simply related as $C_{\phi_+ \phi_- \overline{\omega}} =
C_{\phi_+ \phi_-}^{\omega}$.

The OPE expansion \eqref{ope2} together with \eqref{twopoint} imply that
\begin{equation}
\langle \phi_+(x) \phi_-(0) \overline{\omega}(y) \rangle = {C_{\phi_1 \phi_2 \overline{\omega}}  \over |x|^{\Delta_{+} + \Delta_{-} - \Delta_{\omega}}
 |y|^{2 \ {\omega}}} \left( 1 + {x \over y}  \Delta_{\omega} A + {\overline{x} \over \overline{y}} \Delta_{\omega} A
 + {|x|^2\over |y|^2} \Delta_\omega^2 B + \ldots  \right)
\end{equation}
 on the other hand expanding the 3-point function \eqref{threept12omega} around the $x\rightarrow 0$ limit we find
\begin{equation}
\begin{split}
&\langle \phi_+(x) \phi_-(0) \overline{\omega}(y) \rangle = {C_{\phi_+ \phi_- \overline{\omega}}
\over |x|^{\Delta_{+} + \Delta_{-} - \Delta_{\omega}}  |y|^{2 \Delta_{\omega}}} \Big( 1 + {x \over y}
{(\Delta_+ - \Delta_{-} + \Delta_{\omega})\over 2} \cr
&+ {\overline{x} \over \overline{y}}
{(\Delta_+ - \Delta_{-} + \Delta_{\omega})\over 2} + {|x|^2 \over |y|^2}{(\Delta_+ - \Delta_{-} + \Delta_{\omega})^2\over 4}
+\ldots\Big)
\end{split}
\end{equation}
Comparing the two we find that
\begin{equation}
A =  {(\Delta_+ - \Delta_{-} + \Delta_{\omega}) \over 2 \Delta_{\omega}}\qquad,\qquad B = {(\Delta_+ - \Delta_{-} + \Delta_{\omega})^2 \over 4 \Delta_{\omega}^2}
\end{equation}
We remind the reader that the relevant conformal dimensions here are
\begin{equation}
\begin{split}
&\Delta_{\omega}  = {N^2 - 1 \over  N} \left({1 \over N + k} - {1 \over N + k + 1} \right) \approx {\lambda^2 \over  N}\\
&\Delta_{+} = {N^2 - 1 \over N} \left({1 \over N+k} + {1 \over N+1} \right) \approx 1 + \lambda \\
&\Delta_{-} = {N^2 - 1 \over  N} \left({1 \over N+1} - {1 \over N+ k+1}\right) \approx 1 - \lambda
\end{split}
\end{equation}

Using the exact relation between $\psi$ and $\omega$ given in \eqref{defpsi}
we see that we should have
\begin{equation}
C_{\phi_+ \phi_-}^{\psi_L} = C_{\phi_+ \phi_-}^{\psi_R} =
{(\Delta_+ - \Delta_{-} + \Delta_{\omega}) \over 2 \Delta_{\omega}} \sqrt{ \Delta_{\omega}} C_{\phi_+ \phi_-}^\omega
\end{equation}
\begin{equation}
C_{\phi_+ \phi_-}^{\psi} ={(\Delta_+ - \Delta_{-} + \Delta_{\omega})^2 \over 4 \Delta_{\omega}^2} \sqrt{ \Delta_{\omega}} C_{\phi_+ \phi_-}^\omega
\end{equation}
Using the dimensions above, we precisely get
\begin{equation}
C_{\phi_+ \phi_-}^{\psi_L} = C_{\phi_+ \phi_-}^{\psi_R} = \sqrt{N} C_{\phi_+ \phi_-}^\omega
\end{equation}
and
$$
C_{\phi_+ \phi_-}^{\psi} =  N C_{\phi_+ \phi_-}^\omega
$$
Assuming that at large $N$ we have $C_{\phi_+ \phi_-}^{\psi}=1$, since $\psi$ plays the role of the double trace operator $\colon\phi_+ \phi_-\colon$,
we find that the other 3-point functions must have precisely the values \eqref{opecoeffs} that we determined from the 4-point function
above.

\subsection{Other three point functions}
We would now like to show that the scalar fields $\phi_{+}$, $\phi_{-}$ and
their multi-trace operators couple not just to $\omega$ but also to other
additional sector fields. Let us invent some notation.
We denote:
\begin{equation}
\label{defomegaprime}
\begin{split}
&\omega'_{-} = \coset[f,asym] \\
&\omega'_{+} = \coset[asym,f] \\
\end{split}
\end{equation}
We discuss these fields in some detail in the appendix, but these are evidently fields from the additional-sector.
The quickest way to see this is through the dimension:
\begin{equation}
\label{omegapmdim}
\Delta_{\omega'_{\pm}} \approx \Delta_{\pm}.
\end{equation}
Since, except for the scalars themselves, there are no fields in the ordinary sector which have the same dimension as $\Delta_{\pm}$, the operators $\omega'_{\pm}$ must belong to the additional sector.

When $x \rightarrow 0$, these fields and their descendants mediate the interactions in $G_{\phi_{-}^2 \phi_{+}}$ and
$G_{\phi_{+}^2 \phi_{-}}$.
However, before we extract the three point functions, we need some information about the representations built on $\omega'_{\pm}$.
We discuss $\omega_{-}'$ and state the results for $\omega_{+}'$ at the end. Using the branching
function \eqref{brfn}, we can check that the representation $\coset[f,asym]$ has two level-1 {\em chiral}
descendants: a conformal descendant and a ${\cal W}_N$-descendant. When we join the two chiral sectors together, we find
that there are four relevant conformal primaries.
\begin{equation}
\omega'_{-}\,, \,\zeta_{L-}\,,\, \zeta_{R-}\,,\, \zeta_{-}
\end{equation}
with
\begin{equation}
\Delta_{\zeta_{L-}} = \Delta_{\zeta_{R-}} = \Delta_{\omega'_{-}} + 1 = \Delta_{\zeta} - 1.
\end{equation}
Below, we normalize  $\omega'_{-}\,, \,\zeta_{L-}\,,\, \zeta_{R-}$ to have a two point function $1$, but
$\zeta_{-}$ to have a two point function $2$. This is because we would like to identify $\zeta_{-}$ with the
triple trace operator $\zeta_{-} \sim \colon \phi_{+} \phi_{-}^2$ (as the OPE coefficients below show) and
so we choose its normalization according to \eqref{normdouble}

If we take the point $x$ close to $0$, we can do an OPE expansion as above:
\begin{equation}
\begin{split}
 \overline{\phi}_{+}(x, \overline{x}) \overline{\phi}_{-}^2(0,0)
 &\sim |x|^{{4\over N}-2}\Big[
 \overline{C}_{\phi_{+} \phi_{-}^2}^{\omega'_-} \left(\overline{\omega}'_- \, (0) + \ldots \right) +
 \overline{C}_{\phi_{+} \phi_{-}^2}^{\zeta_{R-}} \overline{x}\left(\overline{\zeta}_{R-}(0) + \ldots \right)
 \\ & +  \overline{C}_{\phi_{+} \phi_{-}^2}^{\zeta_{L-}} x
 \left(\overline{\zeta}_{L-}(0) + \ldots \right)
 + \overline{C}_{\phi_{+} \phi_{-}^2}^{\zeta_-} |x|^2\left(\overline{\zeta}_-(0)+\ldots\right) +
  \ldots\Big]
\end{split}
\end{equation}
To read off the exact three point functions, we need to start with the most singular term in $x$, then subtract
off the contribution of its descendants to disentangle the contribution of other primary operators. Defining
the auxiliary quantities
\be
c^{\omega'_{-}} = \sqrt{2 \over N}, \quad c^{\zeta_{-}} = \sqrt{\frac{(N-2) (N+1) (N-p-1) (p+3)}{N \left((p-1) N^2-p (p+3)
   N+(p-1)^2\right)}}
\ee
we find that
\be
\begin{split}
&C_{\phi_+, \phi_{-}^2}^{\omega'_{-}} =  \sqrt{2} (c^{\omega'_{-}})^2, \\
&C_{\phi_+, \phi_{-}^2}^{\zeta_{L-}} = C_{\phi_+, \phi_{-}^2}^{\zeta_{R-}} =  \sqrt{2} c^{\omega'_{-}} c^{\zeta_{-}} \\
&C_{\phi_+, \phi_{-}^2}^{\zeta_{-}} \approx (c^{\zeta_{-}})^2.
\end{split}
\ee
If we just want the leading order answer at $N$, we just need to note that contribution from $\omega'_{-}$
and $\psi$ is at different orders in ${1 \over N}$,\footnote{Note that in this case, unlike the case involving $\omega$ above, there are no factors of $N$ between three point functions of operators belonging
to the same conformal family. However, what these three point functions show is that there are factors of $N$ within the same coset representation.} so we directly find:
\begin{equation}
\label{omegamcouplings}
\begin{split}
&C_{\phi_+, \phi_{-}^2}^{\omega'_{-}} \approx {2\sqrt{2} \over N}, \\
&C_{\phi_+, \phi_{-}^2}^{\zeta_{L-}} = C_{\phi_+, \phi_{-}^2}^{\zeta_{R-}} \approx  {2 \over \sqrt{N}}, \\
&C_{\phi_+, \phi_{-}^2}^{\zeta_{-}} \approx 1.
\end{split}
\end{equation}
We can perform a similar exercise with the correlator $G_{\phi_{+}^2 \phi_{-}}$. Here, we just quote the large $N$
answer:
\begin{equation}
\label{omegapcouplings}
\begin{split}
&C_{\phi_-, \phi_{+}^2}^{\omega'_{+}} \approx {2\sqrt{2} \over N}, \\
&C_{\phi_-, \phi_{+}^2}^{\zeta_{L+}} = C_{\phi_+, \phi_{+}^2}^{\zeta_{R+}} \approx  {2 \over \sqrt{N}}, \\
&C_{\phi_-, \phi_{+}^2}^{\zeta_{+}} \approx 1.
\end{split}
\ee

%%% Local Variables:
%%% mode: latex
%%% TeX-master: "wncorrelators_paper"
%%% End:

\section{Conclusion and Discussion \label{sec:interpret}}
The four point correlators given in section \ref{list4pt} and the three point functions in \eqref{opecoeffs} and
in \eqref{omegamcouplings} and \eqref{omegapcouplings} are the main results of this paper. In this section we explore the consequences of these results.

First, these correlators show that the ${\cal W}_N$ minimal model has a good large $N$ expansion. Even though there are $\sim e^{ \sqrt{N}}$ additional-sector fields at the same energy as $\phi_{+}$ and $\phi_{-}$, the four point functions \eqref{firstcorr} and \eqref{secondcorr} show that these are both
generalized free fields with double-trace states that have anomalous dimensions of ${\cal O}\left({1 \over N}\right)$.

We now turn to \eqref{mixed4pt} which leads to \eqref{opecoeffs}. Equation \eqref{opecoeffs} shows that
 $\omega$ mediates an interaction between $\phi_{+}$ and $\phi_{-}$. The surprising conclusion that follows from these three point functions is that without the inclusion of additional light fields in the bulk, the bulk theory
will not correctly reproduce boundary correlators even at {\em tree level}.
This is because the primary operators $\psi_L$ and
$\psi_R$, which are descendants of $\omega$,  provide a ${1 \over N}$ correction to the connected four point function of $\phi_{+}$ and $\phi_{-}$. This is the same order as the leading correction from other sources such as the anomalous
dimension of the double trace operator $\psi$, which is also ${\cal O}\left({1 \over N}\right)$.

We should emphasize that this cannot be fixed by tinkering with the coupling constant in a contact interaction between
$\phi_{+}$ and $\phi_{-}$: no contact interaction can reproduce the correlation function \eqref{mixed4pt}. This is because
we have
\begin{equation}
\label{smallxlargeN}
G_{\phi_{+} \phi_{-}} \underset{x \rightarrow 0}{\longrightarrow} {1 \over N} ({1 \over x} + {1 \over \bar{x}}) + {\cal O}\left({1 \over N}\right)
\end{equation}
This small $x$ behaviour cannot be reproduced by a contact Witten diagram \cite{Heemskerk:2009pn, Heemskerk:2010ty}.
We provide a quick proof of this fact here. The contact Witten diagram yields a contribution to the connected correlator that can be written as\footnote{In our case, since $\phi_{-}$ is quantized to have dimension $1 - \lambda$, its correlation function must be computed by first computing
the correlator for a field with dimension $1 + \lambda$ and then Legendre transforming it \cite{Klebanov:1999tb}. This does not change the small $x$ behaviour described below.}
 \begin{equation}
G^{\rm bulk}_{\phi_{+} \phi_{-}} = {{\cal N} \over N} \int K_{\Delta_{-}} (\infty, \vec{Y})\, K_{\Delta_{+}}(1,\vec{Y}) \, K_{\Delta_{+}}(x,\vec{Y})\, K_{\Delta_{-}}(0,\vec{Y}) {d^3 \vec{Y} \over Y_z^3},
\end{equation}
where $K$ is a bulk to boundary propagator for a field of dimension $\Delta$\cite{Witten:1998qj,Gubser:1998bc} and $\vec{Y}$ is a point in the interior
of AdS with radial coordinate $Y_z$ and ${\cal N}$ is a numerical constant
that tells us about the strength of the bulk interaction. (This notation is standard, but the reader may
consult \cite{Fitzpatrick:2011ia} for details.) The quickest way to evaluate the behaviour of this function as $x \rightarrow 0$ is by
means of a Mellin integral transform \cite{Mack:2009mi,Mack:2009gy,Penedones:2010ue,Fitzpatrick:2011ia,Paulos:2011ie}. We have
\begin{equation}
 \label{mellin4pt}
 G^{\rm bulk}_{\phi_{+} \phi_{-}}  = {\cal N}' \int_{-i \infty + \epsilon}^{i \infty +  \epsilon} d s \, \int_{-i \infty + \epsilon}^{i \infty + \epsilon} d t \, \Big[x^{-2 s} (1 - x)^{-2 t} \Gamma(s)^2 \Gamma(t) \Gamma(\Delta_{-} - \Delta_{+} + t) \Gamma(\Delta_{+} - s - t)^2 \Big]
 \end{equation}
We see immediately that the double poles of the function $\Gamma(s)^2$ lead to terms in  $G^{\rm bulk}_{\phi_{+} \phi_{-}}$ that scale like $\log(x)$ for small
$x$ but there is no power law singularity at $x = 0$ in $G^{\rm bulk}$.
This is just the result that the contact interaction can only capture the exchange of ``double trace'' operators and cannot reproduce the small $x$ behaviour \eqref{smallxlargeN}. It is easy to verify that adding derivatives
does not change this result.

Therefore, in the bulk
to reproduce the correct four point function \eqref{mixed4pt}, apart from contact interactions between $\phi_{+}$ and $\phi_{-}$ and the tree-level exchange of higher spin fields, we also need a tree-level exchange of $\omega$.
It is far from clear to us that this light field can be introduced and given interactions in some consistent manner
in the Vasiliev theory. What is somewhat encouraging is that the four-point function of $\omega$ (which can easily be
computed using the methods explained in the Appendix) also has a good ${1 \over N}$ expansion. This indicates that it
may be possible to add $\omega$ as a weakly coupled field in the bulk.

We now indicate some constraints that any such programme must obey. The field $\omega$ itself
can presumably be realized through a field of mass
\begin{equation}
m_{\omega}^2  \approx -{2 \lambda^2 \over  N}
\end{equation}
Note that $m_{\omega}^2 \approx -2\Delta_{\omega}$.
This mass is above the Breitenlohner-Freedman bound \cite{Breitenlohner:1982bm} for \ads[3], since  $m_{\omega}^2 > -1$. It admits two possible
quantizations, and we need to choose the one that yields a CFT operator of dimension $\Delta_{\omega}$:
\begin{equation}
\Delta_{\omega} = 1 - \sqrt{1 + m_{\omega}^2}
\end{equation}
Moreover, we know that the field $\omega$ must interact with $\phi_{+}$ and $\phi_{-}$ to get the three point
functions \eqref{opecoeffs}. After working through the normalizations and using the prescription of \cite{Klebanov:1999tb},
we find that if we {\em canonically normalize} the fields corresponding
to the operators $\omega, \phi_{+}, \phi_{-}$ in the bulk\footnote{This is {\em different} from canonically normalizing
the corresponding bulk operators. To compare this result with
\eqref{opecoeffs} we need to account for the normalization of both
the three-point and the two-point functions computed from the bulk \cite{Freedman:1998tz}} then the bulk interaction Lagrangian, to leading order in ${1 \over N}$ must be
\begin{equation}
\label{omegaint}
L_{int} = {4 (1 - \lambda^2) \over \sqrt{2 \pi} N} \omega^b \phi_{+}^b \phi_{-}^b,
\end{equation}
where we have put a superscript $b$ to distinguish the bulk fields from the
boundary operators. Moreover, this interaction Lagrangian leads to an
equation of motion that, near the boundary, mirrors the constraint \eqref{boundaryconstraint}.

We now come to the question of the other additional-sector fields. We have discussed two such fields above which we called $\omega_{+}'$ and $\omega_{-}'$. These fields run in an OPE channel when we compute the correlators \eqref{correlators45} and once again these fields (or more precisely, their coset descendants)
contribute to the leading ${1 \over N}$ term in the four point function. It is
a little more complicated to generalize the bulk construction above to account
for the contribution of these operators. This is because the bulk automatically
generates double trace operators $\colon \omega \phi_{-} \colon$ and we
first need to disentagle their contribution.

We conclude our paper with some open questions. If the construction above
is indeed the correct method of accounting for the contribution of $\omega$
from the bulk, then a very interesting question is whether we can reproduce
all tree-level correlators on the boundary by adding a finite number of
fields to the bulk.   If the answer to this question is in the negative,
and we need to add ${\rm O}(N)$ extra light fields to the bulk to reproduce
even tree-level boundary correlators, then the bulk theory would become very
complicated and possibly intractable.

A second, longer-range question, has to do with quantum corrections in the bulk. So far, in this section, our discussion has been at tree-level in the bulk. However, the correlators
presented in Section \ref{sec:correlators} are {\em exact}. So, for the bulk and the boundary to really be
dual to each other,  a complicated sequence of Witten diagrams in the bulk including
all loops must sum up to give the answers \eqref{firstcorr} and \eqref{secondcorr}. What is the mechanism underlying this? Moreover, how will the bulk reproduce the strict fusion rules \eqref{eq:fusion}. These fusion rules
imply that if we compute a
correlator with only external $\phi_{+}$ fields then even at a high loop order we should not see the propagation of
$\phi_{-}$ or $\omega$ in intermediate channels? From the bulk point of view, this behaviour is rather unusual.

We hope to return to these questions in future work. We also hope that also hope that the results presented in this paper
will stimulate further work in what is an extremely interesting and promising subject.

%%% Local Variables:
%%% mode: latex
%%% TeX-master: "wncorrelators_paper"
%%% End:

\section*{Acknowledgements}
We would like to thank L. Alvarez-Gaume, I. Antoniadis, M. Gaberdiel, R. Gopakumar, T. Hartman, N. Iizuka, E. Kiritsis, W. Lerche, J. Maldacena, S. Minwalla,
M. Rangamani, D. Skliros and Xi Yin for useful discussions. KP would like to thank the University of Barcelona, the organizers of the ``6th Regional String meeting'' in Milos, the organizers of
the Benasque workshop ``Strings 2011'' and the organizers of the Benasque workshop ``Gravity 2011'' for hospitality.
SR is supported by a Ramanujan Fellowship of the Department of Science and Technology of the Government of India.
SR is grateful to CERN and the Institute for Advanced Study (Princeton), where part of this work was completed, for their hospitality. SR would also like to acknowledge the support of the Harvard University Physics Department. KP and SR would like to thank the Perimeter Institute for hospitality during the ``Back to the Bootstrap'' workshop in 2011, where this work was initiated.
\appendix

\section{Coulomb gas representation}
\label{sec:coulombgas}

The diagonal coset CFT $\widehat{su}(N)_k \otimes \widehat{su}(N)_1\over
\widehat{su}(N)_{k+1}$ has a free field realization in terms of $N-1$ free scalars with background
charge \cite{Fateev:1987zh} --- this is the Coulomb gas representation. Notice that for $N=2$ these CFTs become the standard $c<1$ minimal models and the computation of correlators via the Coulomb gas method is reviewed in the textbook by Di Francesco et al. \cite{DiFrancesco:1997nk}, chapter 9,
based on the original papers \cite{Dotsenko:1984nm}, \cite{Dotsenko:1984ad}. Here we review the generalization of that method to the ${\cal W}_N$ minimal models
with $N>2$.

We start with $N-1$ free scalar fields $\phi^i, i=1,...,N-1$. We will use the notation $\repc{\phi}$ to denoted the $N-1$ dimensional vector. Inner products of such vectors are to be understood as $\repc{\phi}\cdot \repc{\psi}\equiv \sum_i \phi^i \psi^i$. We
are working in normalization where the 2-point function of the scalars is
$$
\langle \phi^i(z,\overline{z}) \phi^j(0,0)\rangle = -2 \delta^{ij} \log(|z|^2)
$$
so the normalization of $\phi$ differs by a factor of $\sqrt{2}$ from that in \cite{DiFrancesco:1997nk}.
The energy momentum tensor contains a background charge piece
$$
T(z) = -{1\over 4} \partial\repc{\phi}\cdot \partial\repc{\phi}+ i\repc{\alpha}_0\cdot\partial^2{\repc{\phi}}
$$
with $\repc{\alpha}_0 = \alpha_0  \repc{\rho}$, where $\repc{\rho}$ is the Weyl vector of $SU(N)$ and in order
for the theory to represent the ${\cal W}_N$ CFT we have to choose the constant $\alpha_0$ to be
\begin{equation}
\label{alphazero}
\alpha_0 = \sqrt{1\over 2(k+N)(k+N+1)}
\end{equation}
Let us calculate the central charge. We use the fact that $\repc{\rho}^2 ={N(N^2-1)\over 12} $ and we find
$$
c =(N-1) \left(1-{N(N+1)\over (k+N)(k+N+1)}\right)
$$
which is indeed the central charge of the ${\cal W}_N$ minimal model \eqref{centralcharge}.

The conformal primaries of interest to us are written in the form
$$
V_{\repc{b}} = \,\colon e^{i\repc{b}\cdot\repc{\phi}}\colon
$$
Without the background charge the holomorphic conformal dimension of this operator would be $h_{\repc{b}}=\repc{b}^2$. The background charge shifts the dimension to
\begin{equation}
\label{confdim}
h_{\repc{b}} = \repc{b}\cdot(\repc{b} -2\repc{\alpha}_0)
\end{equation}
and similarly for the antiholomorphic one, so the total conformal dimension is $\Delta_{\repc{b}} = 2\repc{b}\cdot(\repc{b} -2\repc{\alpha}_0)$.

To determine the allowed values of $\repc{b}$, and hence the spectrum of the theory, we have to discuss the conditions for charge conservation. If we consider a correlator
of the form
$$
\langle V_{\repc{b}_1}(z_1)...V_{\repc{b}_n}(z_n)\rangle
$$
in the free theory (i.e. without the background charge term), then charge conservation implies that the correlator is nonzero only if $\repc{b}_1+...+\repc{b}_n = 0$. The background charge modifies this condition to
\begin{equation}
\label{chargecon}
\repc{b}_1+...+\repc{b}_n = 2 \repc{\alpha}_0
\end{equation}
This already shows that there will be some complications in defining correlators,
for example the naive 2-point function of an operator $\langle V_{\repc{b}} V_{-\repc{b}}\rangle$ would be zero.

In order to understand how the Coulomb gas method works in more detail we would have to review
the underlying BRST-type mathematical structure which was formulated by Felder in \cite{Felder:1988zp}, and is also
reviewed in \cite{DiFrancesco:1997nk}\footnote{We would like to thank L. Alvarez-Gaume and W. Lerche for discussions on the Coulomb gas method.}. However this would go beyond our scope so we will continue with a more
practical point of view and simply compute the relevant correlators. The main intuition is that
the coset CFT is a BRST-invariant sector of the full
Hilbert space corresponding to the free bosons (with background
charge).  Primaries of the coset CFT may have more than one
representatives in the free boson CFT. For example we identify the operator $V_{\repc{b}}$ with $V_{2\repc{\alpha}_0-\repc{b}^*}$, where $\repc{b}^*$ denotes the conjugate representation.

An important point is that when computing correlation functions we can insert ``screening operators'', i.e. operators of the form
$$
Q= \oint dz V_{\repc{b}}(z)
$$
These are nonlocal operators. If we choose $\repc{b}$ so that $V_{\repc{b}}$ has holomorphic dimension $h=1$ then the integrated correlator has (formally) dimension $0$. These operators carry charge and can be inserted in order to saturate the charge conservation condition \eqref{chargecon}.

What are the conditions on $\repc{b}$ in order for $V$ to have holomorphic dimension $1$? It turns
out that we can take
$$
\repc{b} = \alpha \repc{e}_i
$$
where $\repc{e}_i$ is one of the positive roots  of $SU(N)$ and $\alpha$ some constant
which we will determine now. Plugging into the formula \eqref{confdim} for the conformal dimension we find $h = 2\alpha^2 -2 \alpha_0 \alpha $. Solving $h=1$ we find two solutions for $\alpha$ given by
\begin{equation}
\label{apmconst}
\alpha_{\pm} = {1\over 2} (\alpha_0 \pm
\sqrt{\alpha_0^2 + 2})
\end{equation}
So we have two classes of charge screening operators, labeled by positive roots of $SU(N)$
$$
Q^+_i = \oint dz V_{\alpha_+ \repc{e}_i}(z)\qquad,\qquad Q^-_i =
\oint dz V_{\alpha_- \repc{e}_i}(z)
$$
Now consider a correlation function in the Coulomb gas CFT
$$
\langle V_{\repc{b}_1}(z_1)...V_{\repc{b}_n}(z_p)\rangle_{CG} \,= \,
\langle V_{\repc{b}_1}(z_1)...V_{\repc{b}_n}(z_p)\prod_{i=1}^{N-1}(Q^+_i)^{m_i} (Q^-_i)^{n_i}\rangle_{free}
$$
where we have to insert various screening operators in order to satisfy the total
charge condition
$$
\sum_{i=1}^p \repc{b}_i + \alpha_+\sum_{i=1}^{N-1} m_i \repc{e}_i
+ \alpha_- \sum_{i=1}^{N-1} n_i \repc{e}_i = 2 \repc{\alpha}_0
$$
For given $\repc{b}$'s the correlator is nonzero if we can find non-negative integers $m_i,n_i$ such that this condition is satisfied.

In particular (assuming the existence of certain null vectors) this fixes the allowed values of $\repc{b}$ i.e. the spectrum of the CFT. The primaries are labeled by two sets of positive integers $l_i,l'_i$ such that
$$
\sum_i l_i \leq k+N,\qquad \sum_i l'_i \leq k+N-1
$$
and then
$$
\repc{b} = \sum_i \left[\alpha_+(1-l_i) + \alpha_- (1-l_i')\right]\repc{\omega}_i
$$
where $\repc{\omega}_i$ are the fundamental weights of $SU(N)$. The total conformal dimension (i.e. $\Delta=h+\overline{h}$) is
$$
\Delta= 2\left[\sum_i (l_i \alpha_+ + l'_i \alpha_-)\repc{\omega}_i\right]^2-2\repc{\alpha}_0^2
$$
The operator corresponding to the ${\cal W}_N$ scalar $\phi_+$ of representation $\coset[f,1]$ is $l_i = (2,1,..1)$ and $l'_i=(1,1..,1)$ so
$$
\coset[f,1]\qquad  \leftrightarrow \qquad \repc{b} = -\alpha_+ \repc{\omega}_1$$
The formula above gives a conformal dimension
$$
\Delta = 2a_+^2 \repc{\omega}_1\cdot\repc{\omega}_1+ 4 \alpha_+ \alpha_0 \repc{\rho}\cdot\repc{\omega}_1
$$
Using that $\repc{\rho} =  \sum_i \repc{\omega}_i$ and the quadratic form for $\repc{\omega_i}\cdot\repc{\omega_j}$ for $SU(N)$ we find
$$
\Delta = 2\alpha_+^2 {N-1\over N} + 2\alpha_+ \alpha_0 {N(N-1)\over N} =
{N-1\over N}\left(1+{N+1\over N+k}\right)
$$
which is indeed correct for the field $\phi_+=\coset[f,1]$ as presented in \eqref{eq:dimprim}.

For the field $\overline{\phi}_+=\coset[\overline{f},1]$ we have $l_i = (1,1,..,2)$ and $
l'_i=(1,1,..,1)$ so
$$
\coset[\overline{f},1]\qquad \leftrightarrow \qquad \repc{b} = -a_+ \repc{\omega}_{N-1}
$$
Again we find the right conformal dimension.

Let us now look at the other scalars.
For $\phi_-=\coset[1,f]$ we have $l_i = (1,1,..1)$ and $l'_i = (2,1,...,1)$ so
$$
\coset[1,f] \qquad \leftrightarrow \qquad \repc{b} = -\alpha_- \repc{\omega}_1
$$
and conformal dimension
$$
\Delta = 2\alpha_-^2 {N-1\over N} + 2\alpha_- \alpha_0 {N(N-1)\over N}
={N-1\over N}\left(1- {N+1\over N+k+1}\right)
$$
which is the right one. And finally for $\overline{\phi}_-=\coset[1,\overline{f}]$ we
have $\repc{b} = -\alpha_- \repc{\omega}_{N-1}$.

\noindent {\bf Summary:} The scalars of the coset correspond
to vertex operators of the form $V_{\repc{b}} = \colon e^{i \repc{b}\cdot\repc{\phi}}\colon$ with
\begin{equation}
\begin{split}
& \phi_+=\coset[f,1] \qquad \leftrightarrow \qquad \repc{b} = -\alpha_+ \rep{\omega_1}\cr
& \overline{\phi}_+=\coset[\overline{f},1] \qquad \leftrightarrow \qquad \repc{b} = -\alpha_+
\rep{\omega_{N-1}}\cr
& \phi_-=\coset[1,f] \qquad \leftrightarrow \qquad \repc{b} = -\alpha_- \rep{\omega_1}\cr
& \overline{\phi}_-=\coset[1,\overline{f}] \qquad \leftrightarrow \qquad \repc{b} = -\alpha_-\rep{
\omega_{N-1}}
\end{split}
\end{equation}
where $\repc{\omega_i}$ are the fundamental weights of $SU(N)$ and the constants $\alpha_\pm$ are given
via equations \eqref{alphazero} and \eqref{apmconst}.

Now we proceed with the computation of correlation functions. The desired 4-point function are of the form
$$
\langle V_{\repc{b}_1}(z_1)
V_{\repc{b}_2}(z_2)V_{\repc{b}_3}(z_3) V_{\repc{b}_4}(z_4) \rangle_{CG}
$$
which can be computed by a free-field correlator with the insertions
of appropriate screening charges to saturate the total charge
$$
\langle V_{\repc{b}_1}(z_1)
V_{\repc{b}_2}(z_2)V_{\repc{b}_3}(z_3) V_{\repc{b}_4}(z_4) \rangle_{CG}=
\langle V_{\repc{b}_1}(z_1)
V_{\repc{b}_2}(z_2)V_{\repc{b}_3}(z_3) V_{\repc{b}_4}(z_4) \prod_i
Q^+_i \prod_j Q^-_j\rangle_{free}
$$
or
\begin{equation}
\label{finalfreecg}
\begin{split}
\langle V_{\repc{b}_1}(z_1) &
V_{\repc{b}_2}(z_2)V_{\repc{b}_3}(z_3) V_{\repc{b}_4}(z_4) \rangle_{CG}=\cr
&=\prod_i \oint du_i \prod_j \oint dv_j
\,\langle V_{\repc{b}_1}(z_1)
V_{\repc{b}_2}(z_2)V_{\repc{b}_3}(z_3) V_{\repc{b}_4}(z_4) V_{i_1}^+(u_1)...
V_{j_1}^-(v_1)...\rangle_{free}
\end{split}
\end{equation}
where $V^\pm$ are the operators of (total) conformal dimension 2 with $\repc{b} = \alpha_{\pm}
\repc{e}_i$. The choice of the contour of integration will be discussed later and notice that we have only
written down explicitly the holomorphic integration, a similar integration is implied in the antiholomorphic sector.

Notice that the free field correlator which appears above has a very
simple form
\begin{equation}
\label{freefcor}
\langle V_{\repc{b}_1}(z_1)...V_{\repc{b}_n}(z_n)\rangle_{free} = \prod_{i<j}
|z_i - z_j|^{4 \repc{b}_i\cdot\repc{b}_j}
\end{equation}
so the difficulty is in performing the contour integrations mentioned
above.

First we will try to compute the 4-point function
$$
G_{\phi_+\phi_+}(x) = \langle \phi_+(\infty) \overline{\phi}_+(1)\phi_+(x) \overline{\phi}_+(0)\rangle
$$
where $\phi_+ = (\rep{f},\rep{0})$. According to the previous discussion we have
$$
G_{\phi_+\phi_+}(x) = \langle V_{-\alpha_+\repc{\omega}_1}(\infty)
V_{-\alpha_+ \repc{\omega}_{N-1}}(1)
V_{-\alpha_+ \repc{\omega}_1}(x)
V_{-\alpha_+ \repc{\omega}_{N-1}}(0)\rangle_{CG}
$$
but we are also able to switch each of these operators with their equivalent pictures
$V_{\repc{b}}\sim V_{2\repc{\alpha}_0 - \repc{b}^*}$. We are free to choose the right picture for each operator so that the minimum number of screening charges will have to be inserted to saturate the charge conservation condition.

Let us change the picture of the first operator to
$$
G_{\phi_+\phi_+}(x) = \langle V_{2\repc{\alpha}_0+\alpha_+\repc{\omega}_{N-1}}(\infty)
V_{-\alpha_+ \repc{\omega}_{N-1}}(1)
V_{-\alpha_+ \repc{\omega}_1}(x)
V_{-\alpha_+ \repc{\omega}_{N-1}}(0)\rangle_{CG}
$$
We impose the charge neutrality to find what kind of screening operators we have to insert
$$
-\alpha_+ \repc{\omega}_1 -\alpha_+\repc{\omega}_{N-1}+ 2\repc{\alpha}_0  + ({\rm screening}) =
2\repc{\alpha}_0
$$
which means
$$
({\rm screening}) = \alpha_+ (\repc{\omega}_1+\repc{\omega}_{N-1})
$$
or
\begin{equation}
\label{screeninops1}
({\rm screening}) = \alpha_+ (\repc{e}_1+...+\repc{e}_{N-1})
\end{equation}
so we have to insert a screening operators of the $Q^+$ type for each of the positive simple roots of $SU(N)$.

\subsection{Computation of the first 4-point function}

We are interested in computing the simple four point function $G_{\phi_+ \phi_+}$. We will bring the point at $\infty$ to $w$ in intermediate steps to make the presentation more symmetric. In the end we will send $w\rightarrow \infty$ again.

Using the inner products $\repc{\omega}_i\cdot\repc{e}_j=\delta_{ij}$ and $\repc{\omega}_1^2 = \repc{\omega}_{N-1}^2= {N-1\over N}\,,\, \repc{\omega}_1\cdot\repc{\omega}_{N-1}= {1\over N}$ we find that the correlator to be integrated is
\begin{equation}
\begin{split}
G_{\phi_+\phi_+}(x) &= \langle V_{2\repc{\alpha}_0+\alpha_+\repc{\omega}_{N-1}}(w)
V_{-\alpha_+ \repc{\omega}_{N-1}}(1)
V_{-\alpha_+ \repc{\omega}_1}(x)
V_{-\alpha_+ \repc{\omega}_{N-1}}(0)\rangle_{CG} \\
&= \oint dt_1 ...\oint dt_{N-1} (w-x)^{\gamma_x} (w - 1)^{\gamma_1} w^{\gamma_0} \\ &\times x^{2\alpha_+^2 {1 \over N}} (1-x)^{2\alpha_+^2 {1\over N}} (t_1-x)^{-2\alpha_+^2} (t_{N-1}-1)^{-2\alpha_+^2} (t_{N-1})^{-2\alpha_+^2} (t_{N-1} - w)^{2 \alpha_+^2 + 4 \alpha_0 \alpha_+}  \\
&\times \prod_{i=1}^{N-2}(t_i-t_{i+1})^{-2\alpha_+^2} (t_i - w)^{4 \alpha_+ \alpha_0}\quad\times \quad{\rm antiholomorphic}
\end{split}
\end{equation}
Note that this integral is conformally invariant since we can check that the
total power of each $t_i$ is exactly $-2$, since each $t_i$ comes with the
``total exponent''  that is $-4 \alpha_+^2 + 4 \alpha_0 \alpha_+ = -2$. Here $\gamma_x, \gamma_0, \gamma_1$
are some numbers that can easily be computed but will not be of interest to us.

It would seem that this integral involves $N-1$ integrals. However, we
actually only have to do $1$. This is because the value of the integrals
over $t_1, \ldots t_{N-2}$ is fixed by conformal invariance. Notice that each
of these integrals is of the form (focusing only on the holomorphic part)
\begin{equation}
\label{conformalintegral}
I_{\cal C} = \prod_{j=1}^3 \int (t_i - w_j)^{2 \beta_j},
\end{equation}
with $\sum \beta_j = 1$.
The value of the integral in \eqref{conformalintegral} is given by
\begin{equation}
I_{\cal C} \propto (w_1 - w_2)^{2 (\beta_1 + \beta_2 - \beta_3)} (w_2 - w_3)^{2 (\beta_2 + \beta_3 - \beta_1)} (w_3 - w_1)^{2 (\beta_3 + \beta_1 - \beta_2)},
\end{equation}
where we have not fixed the overall constant which is independent of $w_i$.

In fact, we can do all the integrals from $t_1, \ldots t_{N-2}$ at
one shot using conformal invariance. We have
\begin{equation}
\begin{split}
&\int dt_1 \ldots t_{N-2} (t_1 - x)^{-2 \alpha_+^2} \prod_{i=1}^{N-2}(t_i-t_{i+1})^{-2\alpha_+^2} (t_i - w)^{4 \alpha_+ \alpha_0} \\ &= (t_{N-1} - x)^{-2 \alpha_+^2 - 2 \alpha_+ \alpha_0 (N-2)} (t_{N-1} - w)^{2 \alpha_+ \alpha_0 (N-2)} (x - w)^{2 \alpha_+ \alpha_0 (N-2)}
\\
&\sim (t_{N-1} - x)^{-2 \alpha_+^2 - 2 \alpha_+ \alpha_0 (N-2)}
\end{split}
\end{equation}
where in the last line we have again thrown away the factors of $w$, which we are going to take to infinity in any case.

This leaves us with the following integral
\begin{equation}
I = \int d t_{N-1} x^{2\alpha_+^2 {N-1 \over N}} (1-x)^{2\alpha_+^2 {1\over N}}  (t_{N-1}-1)^{-2\alpha_+^2} (t_{N-1})^{-2\alpha_+^2} (t_{N-1} - x)^{-2 \alpha_+^2 - 2 \alpha_+ \alpha_0 (N-2)}
\end{equation}
This integral is exactly of the form that is done in Chapter 9 of \cite{DiFrancesco:1997nk}, for the simpler case
of the $c<1$ minimal models, with
the following constants
\begin{equation}
\label{abcfirst}
\begin{split}
&a = -2\alpha_+^2 \\
&b = -2\alpha_+^2\\
&c = -2 \alpha_+^2 - 2 \alpha_+ \alpha_0 (N-2) \\
\end{split}
\end{equation}
We now need to compare this with our previous result. The conditions that we have to impose i.e. correct
monodromy of the correlator, are the same as the problem studied in \cite{DiFrancesco:1997nk} so we can just
plug in the values of $a,b,c$ in the result from Equation (9.88) of that book.

\subsubsection{Simplification of the Answer}
The answer from Di Francesco et al. (9.88) gives us the answer for the four point
correlator up to normalization. To fix the normalization, we need to set
the series expansion of the function near $x \sim 0$ to be
\begin{equation}
G_{\phi_+ \phi_+} = |x|^{-2 \Delta_{+}} + \ldots,
\end{equation}
where $\ldots$ denote higher order terms.

Rescaling the answer to achieve this, we find that the answer for the correlator
is:
\begin{equation}
\begin{split}
&G_{\phi_+ \phi_+} =
|1-x|^{2\frac{k+N+1}{N^2+k N}} |x|^{-2 \Delta_{+}}   \Bigg[ \frac{\csc \big(\frac{(k+2 N-1) \pi }{k+N}\big) \sin \big(\frac{(3 k+4 N+1) \pi
   }{k+N}\big)}{\Gamma \big(\frac{1-N}{k+N}\big)^2
   \Gamma \big(\frac{2 k+3 N}{k+N}\big)^2} \\ &\times \left|x^{\frac{
   k+2 N}{k+N}} \Gamma
   \big(\frac{-N}{k+N}\big) \Gamma \big(\frac{2 k+3
   N+1}{k+N}\big) \, _2F_1\big(\frac{k+2 N-1}{k+N},\frac{2 k+3
   N+1}{k+N};\frac{2 k+3 N}{k+N};x\big)\right|^2        \\ & +
   \left|_2F_1\big(\frac{k+N+1}{k+N},-\frac{1}{k+N};-\frac{N}{k+N};x\big
   )\right|^2 \,
       \big)\Bigg].
\end{split}
\end{equation}
In this form, it is already clear that the correlator has the right series expansion near $x \sim 0$ because the leading singular term comes from the second
Hypergeometric function, and this is just $1$ at $x = 0$.

However, we can write this in a slightly more symmetric form using what is
called Euler's identity:
\begin{equation}
_2F_1(a,b;c;x) = (1 - x)^{c - a - b} F(c-a, c-b;c;x)
\end{equation}
Applying this identity to the first hypergeometric function, we find that
\begin{equation}
\begin{split}
G_{\phi_+ \phi_+} = |(1-x)x|^{-2 \Delta_{+}} \Bigg[ &\left|(1-x)^{k+2N \over k+N}\,  _2F_1\big(\frac{k+N+1}{k+N},-\frac{1}{k+N};-\frac{N}{k+N};x\big)\right|^2 \\ &+ {\cal N}_1 \left|x^{k+2N \over k+N}\,  _2F_1\big(\frac{k+N+1}{k+N},-\frac{1}{k+N};\frac{2k + 3 N}{k+N};x\big)\right|^2 \Bigg],
\end{split}
\end{equation}
with
\begin{equation}
\begin{split}
{\cal N}_1 &= \frac{\csc \big(\frac{(k+2 N-1) \pi }{k+N}\big) \sin \big(\frac{(3 k+4 N+1) \pi
   }{k+N}\big)}{\Gamma \big(\frac{1-N}{k+N}\big)^2
   \Gamma \big(\frac{2 k+3 N}{k+N}\big)^2} \Gamma
   \big(\frac{-N}{k+N}\big)^2 \Gamma \big(\frac{2 k+3
   N+1}{k+N}\big)^2 \\
&=  -{\Gamma({k + 2 N - 1 \over k + N}) \Gamma
   \big(\frac{-N}{k+N}\big)^2 \Gamma \big(\frac{2 k+3
   N+1}{k+N}\big) \over \Gamma({-k - 2N - 1 \over k+N}) \Gamma \big(\frac{1-N}{k+N}\big) \Gamma \big(\frac{2 k+3 N}{k+N}\big)^2
   \big)}
\end{split}
\end{equation}

\subsection{Computation of the second 4-point function}
The formulas here are identical to \eqref{abcfirst}, with the replacement $\alpha_{+} \rightarrow \alpha_{-}$.

\begin{equation}
\begin{split}
&G_{\phi_{-} \phi_{-}} = |1-x|^{2\frac{k+N}{N^2+k N+N}} |x|^{-2 \Delta_{-}}
    \Bigg[\frac{ \csc \big(\frac{(k+2) \pi }{k+N+1}\big) \sin \big(\frac{(3 k+2 N+2) \pi
   }{k+N+1}\big)}{\Gamma
   \big(\frac{N-1}{k+N+1}\big)^2 \Gamma \big(\frac{2 k+N+2}{k+N+1}\big)^2}\\
   & \times \Gamma
   \big(\frac{N}{k+N+1}\big)^2 \Gamma \big(\frac{2
   k+N+1}{k+N+1}\big)^2  \left| x^{\frac{1+k}{k+N+1}} \, _2F_1\big(\frac{k+2}{k+N+1},\frac{2
   k+N+1}{k+N+1}; \frac{2
   k+N+2}{k+N+1}; x)\right|^2  \\ & +  \left|  \,
   _2F_1\big(\frac{k+N}{k+N+1},\frac{1}{k+N+1};\frac{N}{k+N+1};x\big)\right|^2 \Bigg]
\end{split}
\end{equation}
After applying Euler's identity again, we find the result 
\begin{equation}
\begin{split}
G_{\phi_{-} \phi_{-}} = |(1-x)x|^{-2 \Delta_{-}}  \Bigg[  &\left| (1 - x)^{1+k \over N+k+1} \,
   _2F_1\big(\frac{k+N}{k+N+1},\frac{1}{k+N+1};\frac{N}{k+N+1};x\big)\right|^2 \\ &+ {\cal N}_2  \left| x^{\frac{1+k}{k+N+1}} \, _2F_1\big(\frac{k+2}{k+N+1},\frac{2
   k+N+1}{k+N+1}; \frac{2
   k+N+2}{k+N+1};x \big) \right|^2  \Bigg]
\end{split}
\end{equation}
with
\begin{equation}
\begin{split}
{\cal N}_2 &= \frac{ \csc \big(\frac{(k+2) \pi }{k+N+1}\big) \Gamma
   \big(\frac{N}{k+N+1}\big)^2 \Gamma \big(\frac{2
   k+N+1}{k+N+1}\big)^2  \sin \big(\frac{(3 k+2 N+2) \pi
   }{k+N+1}\big)}{\Gamma
   \big(\frac{N-1}{k+N+1}\big)^2 \Gamma \big(\frac{2 k+N+2}{k+N+1}\big)^2}  \\
&= -{ \Gamma
   \big(\frac{N}{k+N+1}\big)^2 \Gamma \big(\frac{2
   k+N+1}{k+N+1}\big)  \Gamma({k + 2 \over k + N + 1}) \over \Gamma
   \big(\frac{N-1}{k+N+1}\big) \Gamma \big(\frac{2 k+N+2}{k+N+1}\big)^2 \Gamma({-k \over k + N + 1})}
\end{split}
\end{equation}
\subsection{Computation of the third  4-point function}

We now consider the mixed correlator
$$
G_{\phi_+\phi_-} \sim \langle V_{-\alpha_-\repc{\omega}_1}(\infty)
V_{-\alpha_+ \repc{\omega}_1}(1)
V_{-\alpha_+ \repc{\omega}_{N-1}}(x)
V_{-\alpha_- \repc{\omega}_{N-1}}(0)\rangle_{CG}
$$
We change the picture of the first operator and we have
$$
G_{\phi_+\phi_-} \sim \langle V_{2\repc{\alpha}_0+\alpha_-\repc{\omega}_{N-1}}(\infty)
V_{-\alpha_+ \repc{\omega}_1}(1)
V_{-\alpha_+ \repc{\omega}_{N-1}}(x)
V_{-\alpha_- \repc{\omega}_{N-1}}(0)\rangle_{CG}
$$
For charge conservation find that the needed screening operators are exactly the same as before.

In this case we find that the (holomorphic part of the) correlator that has to be integrated is
\begin{equation}
\begin{split}
I= &(w-a)^{\gamma_1} (w-x)^{\gamma_2} w^{\gamma_3} (1-x)^{2\alpha_+^2 {1\over N}} x^{2\alpha_+
\alpha_-{N-1\over N}}(t_1-1)^{-2\alpha_+^2}(t_{N-1}-w)^{4\alpha_0 \alpha_+ + 2 \alpha_+
\alpha_-}\times \cr &
\times (t_{N-1}-x)^{-2\alpha_+^2}(t_{N-1})^{-2\alpha_+\alpha_-}
\prod_{i=1}^{N-2}(t_i-t_{i+1})^{-2\alpha_+^2}(t_i-w)^{4\alpha_0 \alpha_+}
\end{split}\end{equation}
Doing the intermediate integrals as before this leaves us with the remaining integral
over $t_{N-1}$ of the form
$$
I = \int dt_{N-1} (1-x)^{2\alpha_+^2 {1\over N}} x^{2\alpha_+
\alpha_-{N-1\over N}} (t_{N-1}-1)^{-2\alpha_+^2-2\alpha_+ \alpha_0(N-2)}(t_{N-1})^{-2\alpha_+\alpha_-}
(t_{N-1}-x)^{-2\alpha_+^2}
$$
This is of the  form from \cite{DiFrancesco:1997nk}, Equation (9.88),  with
\begin{equation}
\begin{split}
&a = -2\alpha_+\alpha_-= 1 \\
&b =  -2 \alpha_+^2 - 2 \alpha_+ \alpha_0 (N-2)\\
&c = -2\alpha_+^2\\
\end{split}
\end{equation}
Plugging everything in we find that the correlator (normalized correctly) has the following simple form
\begin{equation}
\label{mixed4ptb}
\begin{split}
G_{\phi_{+}\phi_{-}}(x) =
|1 - x|^{-2 \Delta_+} |x|^{2/N} &\times \Big[ 1 +{1\over N} \left({1-x \over x} + {1-\overline{x}\over \overline{x}} \right)
+{1\over N^2}\left|{1-x\over x}\right|^2\Big]
\end{split}
\end{equation}

\subsection{Computation of the Fourth and Fifth Four Point Functions}

We will consider the correlator
\begin{equation}
\label{fourthcorrelator}
G_{\phi_{+} \phi_{-}^2}(x) =  \langle \phi_{-}^2(\infty) \phi_{+}(1) \overline{\phi}_{+}(x) \overline{\phi}_{-}^2(0) \rangle.
\end{equation}
followed by another very similar correlator
\begin{equation}
\label{fifthcorrelator}
G_{\phi_{-} \phi_{+}^2}(x) =  \langle \phi_{+}^2(\infty) \phi_{-}(1) \overline{\phi}_{-}(x) \overline{\phi}_{+}^2(0) \rangle.
\end{equation}
where, as we mentioned in \ref{sec:correlators} we use $\phi_{-}^2$ and $\phi_{+}^2$ to denote the following coset primary fields:
\begin{equation}
\phi_{-}^2 \equiv \coset[1,asym], \quad \phi_{+}^2 \equiv \coset[asym,1]
\end{equation}

First, let us understand the $\coset[1, asym]$ representation. With
$\Lambda_{\text{asym}}$ the highest weight vector corresponding to the symmetric tensor
representation, we find that the dimension of the $\coset[1,asym]$ representation
in the coset is
\begin{equation}
\label{dimasym}
\Delta_{\coset[1,asym]} = {(p \,\Lambda_{\rep{asym}} - \weyl{\rho})^2 - \weyl{\rho}^2 \over  p (p + 1)}  = 2 { (N - 2)(p - N) \over N (p + 1)} \approx 2(1-\lambda).
\end{equation}
This is what suggests that suggests that we should identify this field with the following double
trace operator:
\begin{equation}
\coset[1,asym] \sim : \phi_- \phi_-:
\end{equation}
Our choice of notation was guided by this observation.

Now, let us determine the field that runs in the intermediate channel when
we take $x \rightarrow 0$ in \eqref{fourthcorrelator}. The fusion rules \eqref{eq:fusion} tell us that the only field that appears here is $\coset[f,asym]$. This has dimension
\begin{equation}
\label{dimasymf}
\Delta_{\coset[f,asym]} = {((p+1) \rep{f} - p \rep{\Lambda_{\text{asym}}} + \weyl{\rho})^2 - \weyl{\rho}^2 \over p (p + 1)} = \frac{  p (p+3) N - (p-1)^2 - (p-1) N^2}{N p (p+1)} \approx 1 - \lambda
\ee
Note that, in the 't Hooft limit $\Delta_{\coset[asym,f]} \approx \Delta_{-}$. Since there is no other field of dimension $\Delta_{-}$ in the bulk, except for
$\phi_-$, this immediately tells us that $\Delta_{\coset[f,asym]}$ is a ``hidden field.'' We will use the notation
\begin{equation}
\omega'_{-} \equiv  \coset[asym,f]
\end{equation}
Note that
\begin{equation}
\Delta_{\coset[f,asym]} + 1= \Delta_{+} + 2 \Delta_{-}
\end{equation}
So, one of the level-1 descendant of $\coset[f,asym]$ is the triple trace operator. However, using the branching function \eqref{brfn}, we can check that $\coset[f,asym]$ has two level-1 descendants: a conformal descendant and a $W$-descendant. The triple trace operator needs to be identified with a linear combination of these.

Now, we proceed to evaluate the correlator \eqref{fourthcorrelator}. We start with the Coulomb gas formalism. First, we
want to identify the primary $\coset[1,asym]$. This identification is simply:
\begin{equation}
\coset[1,asym] \leftrightarrow -\alpha_- \repc{\omega_2}.
\end{equation}
One check of this identification is the dimension. From the Coulomb gas, we have
\begin{equation}
\begin{split}
\Delta_{\coset[1,asym]} &= 2\alpha_-^2 \repc{\omega_2}^2+ 4 \alpha_- \alpha_0 \weyl{\rho}\cdot\repc{\omega_2} \\ &=2 {k ( N - 2) \over N (1 + k + N)}
\end{split}
\end{equation}
which matches perfectly with \eqref{dimasym}.

The Coulomb gas correlator that we need to compute can be written:
\bea
&\langle V_{-\alpha_{-} \repc{\omega_2}}(w) V_{-\alpha_{+} \repc{\omega_1}}(1) V_{-\alpha_{+} \repc{\omega_{N-1}}}(x) V_{-\alpha_{-} \repc{\omega_{N-2}}}(0) \rangle_{CG} \\ &=  \langle V_{\repc{\alpha_0} + \alpha_{-} \repc{\omega_{N-2}}}(w) V_{-\alpha_{+} \repc{\omega_1}}(1) V_{-\alpha_{+} \repc{\omega_{N-1}}}(x) V_{-\alpha_{-} \repc{\omega_{N-2}}}(0) \rangle_{CG}
\eea
Hence, the screening operators are the same as \eqref{screeninops1}. Note that in the end we will take $w \rightarrow \infty$.

This leads to the following integrand.
\begin{equation}
\label{integrand}
\begin{split}
&G_{\phi_{-}^2 \phi_{+}} = (w - x)^{\gamma_x} (w - 1)^{\gamma_1} w^{\gamma_0}
(1 - x)^{2 \alpha_+^2 \over N} x^{2 \alpha_{+} \alpha_{-} (N-2) \over N} \oint dt_1 ...\oint dt_{N-1} \\ &\Big[
(t_1 - 1)^{-2 \alpha_{+}^2} (t_{N-1} - x)^{-2 \alpha_{+}^2} (t_{N-2})^{-2 \alpha_{-} \alpha_{+}} (t_{N-2} - w)^{2 \alpha_{-} \alpha_{+}} \prod_{i=1}^{N-1} (t_i - w)^{4 \alpha_{+} \alpha_{0}} (t_i - t_{i+1})^{-2 \alpha_+^2} \Big]
\end{split}
\end{equation}
Let us first do the integrals from $t_1 \ldots t_{N-3}$. We have
\begin{equation}
\label{firstnminus3}
\begin{split}
&\oint dt_1 ...\oint dt_{N-3}(t_1 - 1)^{-2 \alpha_{+}^2} \prod_{i=1}^{N-3} (t_i - w)^{4 \alpha_{+} \alpha_{0}} (t_i - t_{i+1})^{-2 \alpha_+^2}  \\
&\propto (w - 1)^{2 (N-3) \alpha_+ \alpha_0} (t_{N-2} - 1)^{-2 \alpha_{+}^2 - 2 (N - 3) \alpha_{+} \alpha_0} (t_{N-2}-w)^{2 (N-3) \alpha_{+} \alpha_0}.
\end{split}
\end{equation}

We can also separately do the integral over $t_{N-1}$.
\begin{equation}
\label{lastinteg}
\begin{split}
&\int d t_{N-1} (t_{N-1} - x)^{-2 \alpha_+^2} (t_{N-1} - w)^{4 \alpha_{+} \alpha_{0}} (t_{N-1} - t_{N-2})^{-2 \alpha_+^2}
\\
&\propto (w - x)^{2 \alpha_{+} \alpha_0} (w - t_{N-2})^{2 \alpha_{+} \alpha_0} (t_{N-2} - x)^{-2 \alpha_+^2 - 2 \alpha_{+} \alpha_0}
\end{split}
\end{equation}

We now substitute \eqref{firstnminus3} and \eqref{lastinteg} into \eqref{integrand} and take $w \rightarrow \infty$. This
leaves us with
\begin{equation}
G_{\phi_{-}^2 \phi_{+}} \propto (1 - x)^{2 \alpha_+^2 \over N} x^{2 \alpha_{+} \alpha_{-} (N-2) \over N} \oint_{t_{N-2}} t_{N-2}^{-2 \alpha_{-} \alpha_{+}}  (t_{N-2} - 1)^{-2 \alpha_{+}^2 - 2 (N-3)\alpha_{+} \alpha_{0}} (t_{N-2} - x)^{-2 \alpha_+^2 - 2 \alpha_{+} \alpha_{0}}.
\end{equation}
This is an integral of the Di Francesco form, with
\begin{equation}
\label{abcfor4}
\begin{split}
a &= -2 \alpha_{-} \alpha_{+} \\
b &= -2 \alpha_{+}^2 - 2 (N-3)\alpha_{+} \alpha_{0}\\
c &= -2 \alpha_+^2 - 2 \alpha_{+} \alpha_{0} \\
\end{split}
\end{equation}

After using Di Francesco's formula, and normalizing the integral correctly, we find
\begin{equation}
\label{prelimfourth}
G_{\phi_{-}^2 \phi_{+}} = {8 \over N^2} \left| (1-x)^{\frac{k+N+1}{N^2+k N}} x^{\frac{2}{N}-1}
    \,
   _2F_1\left(\frac{k+N+2}{k+N},\frac{N}{k+N};\frac{2}{k+N};x\right) \right|^2
\end{equation}
Actually we can simplify this greatly by applying Euler's identity
\begin{equation}
\begin{split}
 _2F_1\left(\frac{k+N+2}{k+N},\frac{N}{k+N};\frac{2}{k+N};x\right) &=
(1 - x)^{-k - 2 N \over k + N}\, _2F_1\left(-1,\frac{2-N}{k+N};\frac{2}{k+N};x\right) \\
&= (1 - x)^{-k - 2 N \over k + N} \left(1 + {N - 2 \over 2} x\right).
\end{split}
\end{equation}

Even the complicated powers of $(1-x)$ simplify when combines with the
other terms in \eqref{prelimfourth}. Our final answer can be written in the form
\begin{equation}
G_{\phi_{-}^2 \phi_{+}} = {8 \over N^2} \left|(1-x)^{-\Delta_+} x^{\frac{2}{N}-1} \left(1 + {N - 2 \over 2} x\right)\right|^2.
\end{equation}
Note that the number that appears in the exponent of $x$ is also
\begin{equation}
{\Delta_{\coset[f,0]} + \Delta_{\coset[0,asym]} - \Delta_{\coset[f,asym]} \over 2} = {2 \over N} - 1
\end{equation}

We can check that
\begin{equation}
G_{\phi_{-}^2 \phi_{+}} \underset{N \rightarrow \infty}{\longrightarrow} 2|1-x|^{-2 (1 + \lambda)}
\end{equation}
which is exactly what we expect.
Second, we can also check that
\begin{equation}
G_{\phi_{-}^2 \phi_{+}} \underset{x \rightarrow 1}{\longrightarrow}2
|1-x|^{-2 \Delta_+},
\end{equation}
which tells us that the correlator is normalized correctly.

The fifth correlator \eqref{fifthcorrelator} can be computed by just replacing $\alpha_{+} \leftrightarrow \alpha_{-}$
in \eqref{abcfor4}. Simplifying the final answer, we get
\begin{equation}
G_{\phi_{+}^2 \phi_{-}}(x) =  {8 \over N^2} \left|(1-x)^{-\Delta_-} x^{\frac{2}{N}-1} \left(1 + {N - 2 \over 2} x\right)\right|^2.
\end{equation}

%%% Local Variables:
%%% mode: latex
%%% TeX-master: "wncorrelators_paper"
%%% End:

\section{Coset Correlators in terms of WZW Correlators \label{sec:bratchikov}}
In this appendix we describe how correlation functions in the coset theory can be computed in terms of ordinary
correlators in the WZW model. WZW correlators can be computed using the Knizhnik-Zamolodchikov (KZ) equations; in fact,
for the four point functions that we consider in this paper, the WZW correlators that we require are already given
in the standard textbook by Di Francesco et al.\cite{DiFrancesco:1997nk}

The method of reducing coset correlation functions to correlators in the parent WZW models was developed by
Gawedzki and Kupiainen \cite{Gawedzki:1988hq,Gawedzki:1988nj} and then elaborated by Bratchikov \cite{Bratchikov:2000mh}.
The idea of this method is to write the $\hat{g}/\hat{h}$ coset model is to gauge the $H$ subgroup in the $G$-WZW model. By performing
some manipulations in the functional integral, Gawedzki and Kupiainen showed that the path integral over the
gauged model reduced to the product of a path integral over the ordinary $G$-WZW model and a path integral over a
``ghost'' H model. The correlators of this ``ghost'' model obey the same KZ equations as the $H$ model but with
the level $k$ replaced by $k \rightarrow -k - 2 g_H$, where $g_H$ is the dual Coxeter number of $H$. This is the same
as replacing $k + g_H \rightarrow -k - g_H$.

The practical question, as in the Coulomb gas case above, is to identify the correct operators corresponding to
the primaries of the coset in the product of these decoupled WZW models. The procedure to do this was elucidated in a nice paper by Bratchikov.\cite{Bratchikov:2000mh} As we have explained above primaries of the coset are given by specifying
a primary of $\hat{g}$ and a primary of $\hat{h}$. Let us say that we wish to construct the coset primary field
$$
\psi_{RL}
$$
where $R$ is a representation of $G$ and $L$ is a representation of $H$.

The first step is to identify operators in the representation of the
{\em affine algebra} $\hat{g}$ built on $R$ that transform in the representation $L$. In general, these operators may not be affine primaries although in all the cases that we consider in this paper, this complication does not occur. We call
these operators
\be
\label{chigrl}
\chi_{\hat{g},RL}^i
\ee
where $i$ indexes the various operators in the representation $L$ and so $i = 1 \ldots \text{dim}(L)$.

Note that $\chi_{\hat{g},RL}^i$ are all still fields in the $G$-WZW model
and we have indicated the representations they transform in --- $L$ --- and the representations they descended
from --- $R$.

Now, we consider a fictitious $H$-WZW model with level $-k - 2 g_H$, and consider primary fields
\be
\label{chihbarl}
\chi^{\hat{h}^*, \bar{L}}_i
\ee
 that transform in the representation $\bar{L}$. Once again $i = 1 \ldots \text{dim}{L}$, but
we have lowered it to indicate that this field transforms in the representation $\bar{L}$. The the coset
primary field is
identified with the operator
\begin{equation}
\label{cosetprimarybratchikov}
\psi_{R L} = \sum_{i = 1}^{{\rm dim}(L)} \chi_{\hat{g},R L}^{i} \chi^{\hat{h}*,\bar{L}}_{i}.
\end{equation}
Note that computing correlators of \eqref{cosetprimarybratchikov} is just a question of computing correlators
in the parent $G$ model and the ``ghost'' $H$ model and contracting them together.

\subsection{Construction of the Scalar Fields}
We now indicate how this works for the correlators that we are interested in and give explicit
expressions for the coset primaries corresponding to the scalar fields in this formalism. In our case,
\begin{equation}
\widehat{g} = \widehat{su}(N)_k \times \widehat{su}(N)_1, \quad H = \widehat{su}(N)_{k+1}
\end{equation}
However, the procedure above tells us that we need to consider a ``ghost model'' with level
\be
k^* = -k - 2 N - 1
\ee

We will denote the basic fields of the $\widehat{su}(N)$ model by $g_{a}^{\bar{b}}(x,\bar{x})$ which transforms as a fundamental
representation under the left-moving currents and an anti-fundamental representation under the right-moving
currents. Since there are several $\widehat{su}(N)$ models in our case, we place a subscript under $g$ to indicate which model we are referring to. We will use the convention that unbarred (barred) indices transform under left (right) affine transformations, raised indices transform as anti-fundamentals of $SU(N)$, and lowered indices transform as fundamentals of $SU(N)$.

The complex partner of this field is $\bar{g}^{a}_{\bar{b}}(x,\bar{x})$ which transforms as a fundamental
representation under the left-moving currents and an anti-fundamental representation under the right-moving
currents.

The construction of $\phi_{+}$ is the simplest. This is because, for $\phi_{+}$, $L = 0$ (in the notation above). So
we have
\be
\phi_{+}(x, \bar{x}) \leftrightarrow (g_{k})_{a}^{\bar{b}} (x, \bar{x}) (\bar{g}_{1})_{\bar{b}}^{a}(x, \bar{x})
\ee
We do not even need to normal order this product since the operators on the right hand side belong to two decoupled
theories.

The construction of $\phi_{-}$ is a little more complicated. In the notation above, for $\phi_{-}$, we have $R = \rep{f}$
and $L = \rep{f}$. First, we construct the fields corresponding to \eqref{chigrl}
\be
(\chi_{\hat{g},\rep{f}, \rep{f}})_{a}^{\bar{b}} = (g_k)_{a}^{\bar{b}}
\ee
and the field corresponding to \eqref{chihbarl}
\be
(\chi_{\hat{h}^*,\rep{\bar{f}}})_{\bar{b}}^{a} = (\bar{g}_{k^*})_{\bar{b}}^{a}
\ee
This leads to the following expression for the coset primary field $\phi_{-}$
\be
\phi_{-} \leftrightarrow (g_k)_{a}^{\bar{b}} (\bar{g}_{k^*})_{\bar{b}}^{a}
\ee
In subsection \ref{subsec45bratch}, we will need additional vertex operators corresponding to the fields $\phi_{-}^2 \equiv \coset[0,asym]$ and $\phi_{+}^2 \equiv \coset[asym,0]$. These follow quite naturally from the procedure above. We have
\be
\begin{split}
&\phi_{+}^2 \leftrightarrow \sqrt{2} ({\cal A}_k)_{[a_1,a_2]}^{[\bar{b_1}, \bar{b_2}]} ({\cal \bar{A}}_1)^{[a_1,a_2]}_{[\bar{b_1}, \bar{b_2}]}\\
&\phi_{-}^2 \leftrightarrow \sqrt{2} ({\cal A}_1)_{[a_1,a_2]}^{[\bar{b_1}, \bar{b_2}]}  ({\cal \bar{A}}_{k*})^{[a_1,a_2]}_{[\bar{b_1}, \bar{b_2}]}
\end{split}
\ee
where ${\cal A}_{[a_1,a_2]}^{[\bar{b_1}, \bar{b_2}]}$ is the primary that transforms in the anti-symmetric representation
on the left and its conjugate on the right. We comment more on the normalization
in subsection \ref{subsec45bratch}

\subsection{Computation of the first three correlators}
In the Table below, we now collect together the various representations involved in the three correlators in \eqref{firstcorrelator} -- \eqref{thirdcorrelator}. The subscript below the representations below indicates the position at which
this representation is placed in the correlator. These three correlators are particularly easy to compute since
the required four point WZW functions are easily available in the literature.
\begin{equation}
\begin{array}{|c|c|c|c|} \hline
\text{Correlator}&\widehat{su}(N)_k~\text{reps}&\widehat{su}(N)_1~\text{rep}&\widehat{su}(N)_{k+1}~\text{rep}\\ \hline & & &\\
G_{\phi_+ \phi_+}& \langle \rep{f}_{\infty},  \rep{f}_{1}, \rep{\bar{f}}_{x} , \rep{\bar{f}}_0 \rangle & \langle \rep{\bar{f}}_{\infty}, \rep{\bar{f}}_{1},  \rep{f}_{x},  \rep{f}_0 \rangle & \langle \rep{1}_{\infty}, \rep{1}_{1}, \rep{1}_{x}, \rep{1}_{0} \rangle \\
G_{\phi_- \phi_-}& \langle \rep{1}_{\infty}, \rep{1}_{1}, \rep{1}_{x}, \rep{1}_{0} \rangle & \langle \rep{f}_{\infty},  \rep{f}_{1}, \rep{\bar{f}}_{x} , \rep{\bar{f}}_0 \rangle & \langle \rep{\bar{f}}_{\infty}, \rep{\bar{f}}_{1},  \rep{f}_{x},  \rep{f}_0 \rangle \\
G_{\phi_{+} \phi_{-}} & \langle \rep{f}_{\infty},\rep{1}_1, \rep{\bar{f}}_x, \rep{1}_0 \rangle & \langle \rep{\bar{f}}_{\infty}, \rep{\bar{f}}_{1}, \rep{f}_{x}, \rep{f}_0 \rangle & \langle \rep{1}_{\infty}, \rep{f}_1, \rep{1}_x, \rep{\bar{f}}_0 \rangle \\ \hline
\end{array}
\end{equation}

This leads to the following expressions for the first of the three correlators.
\begin{equation}
\begin{split}
G_{\phi_{+}\phi_{+}}(x) = &\langle g^{{\bar{b}}_1}_{a_1}(\infty) {\bar{g}}^{b_2}_{\bar{a}_2}(1)
g^{{\bar{b}_3}}_{a_3}(x) {\bar{g}}^{b_4}_{\bar{a}_4}(0) \rangle_{\widehat{su}(N)_k}
\\
\times
&\langle {\bar{g}}^{\beta_1}_{\bar{\alpha}_1}(\infty) g^{\bar{\beta}_2}_{\alpha_2}(1)
{\bar{g}}^{\beta_3}_{\bar{\alpha}_3}(x) g^{\bar{\beta}_4}_{\alpha_4}(0)  \rangle_{\widehat{su}(N)_1}\cr
\times
& \delta^{\bar{\alpha}_1}_{{\bar{b}_1}} \delta^{\alpha_2}_{b_2} \delta^{\bar{\alpha}_3}_{{\bar{b}_3}} \delta^{\alpha_4}_{b_4}
\delta^{a_1}_{\beta_1}\delta^{\bar{a}_2}_{\bar{\beta}_2}\delta^{a_3}_{\beta_3}\delta^{\bar{a}_4}_{\bar{\beta}_4} \times {1 \over N^4}
\end{split}
\end{equation}
The second correlator is given by
\begin{equation}
\begin{split}
G_{\phi_{+}\phi_{+}}(x) = &\langle g^{{\bar{b}_1}}_{a_1}(\infty) {\bar{g}}^{b_2}_{\bar{a}_2}(1)
g^{{\bar{b}_3}}_{a_3}(x) {\bar{g}}^{b_4}_{\bar{a}_4}(0) \rangle_{\widehat{su}(N)_1}
\cr
\times
&\langle {\bar{g}}^{\beta_1}_{\bar{\alpha}_1}(\infty) g^{\bar{\beta}_2}_{\alpha_2}(1)
{\bar{g}}^{\beta_3}_{\bar{\alpha}_3}(x) g^{\bar{\beta}_4}_{\alpha_4}(0)  \rangle_{\widehat{su}(N)_{k^*}}\cr
\times
& \delta^{\bar{\alpha}_1}_{{\bar{b}_1}} \delta^{\alpha_2}_{b_2} \delta^{\bar{\alpha}_3}_{{\bar{b}_3}} \delta^{\alpha_4}_{b_4}
\delta^{a_1}_{\beta_1}\delta^{\bar{a}_2}_{\bar{\beta}_2}\delta^{a_3}_{\beta_3}\delta^{\bar{a}_4}_{\bar{\beta}_4} \times {1 \over N^4}
\end{split}
\end{equation}
The mixed correlator is given by:
\begin{equation}
\label{mixedcorr}
\begin{split}
G_{\phi_{+} \phi_{-}}(x) &= \langle g^{{\bar{b}_1}}_{a_1}(\infty) {\bar{g}}^{b_2}_{\bar{a}_2}(1) g^{{\bar{b}_3}}_{a_3}(x) {\bar{g}}^{b_4}_{\bar{a}_4}(0) \rangle_{\widehat{su}(N)_1}  \\ &\times \langle {\bar{g}}^{\beta_1}_{\bar{\alpha}_1}(\infty)  g^{\bar{\beta}_4}_{\alpha_4}(0) \rangle_{\widehat{su}(N)_{k^*}} \\ &\times \langle g^{\bar{\beta}_2}_{\alpha_2}(1) {\bar{g}}^{\beta_3}_{\bar{\alpha}_3}(x) \rangle_{\widehat{su}(N)_k} \\ &\times \delta^{\bar{\alpha}_1}_{{\bar{b}_1}} \delta^{\alpha_2}_{b_2} \delta^{\bar{\alpha}_3}_{{\bar{b}_3}} \delta^{\alpha_4}_{b_4} \delta^{a_1}_{\beta_1} \delta^{\bar{a}_2}_{\bar{\beta}_2} \delta^{a_3}_{\beta_3} \delta^{\bar{a}_4}_{\bar{\beta}_4} \times {1 \over N^4}
\end{split}
\end{equation}

The building block in all these three expressions is the four point function
of the fields $g^{b}_{a}(x, \bar{x})$ in the $\widehat{su}(N)_k$ WZW model. This four
point function is computed in the textbook \cite{DiFrancesco:1997nk} and conveniently reviewed in
\cite{Kiritsis:2010xc}. For the reader's convenience,
we reproduce the results below.

The four point function may be written
\begin{equation}
{\cal H}^{\bar{b}_1 b_2 \bar{b}_3 b_4}_{a_1 \bar{a}_2 a_3 \bar{a}_4}(x,\bar x)\equiv\langle \langle g^{{\bar{b}_1}}_{a_1}(\infty) {\bar{g}}^{b_2}_{\bar{a}_2}(1) g^{{\bar{b}_3}}_{a_3}(x) {\bar{g}}^{b_4}_{\bar{a}_4}(0) \rangle_{\widehat{su}(N)_k} =\sum_{i,j=1}^2 I^i \bar I^j~ \Omega_{ij}(x,\bar x)
\end{equation}
where
\begin{equation}
I^1=\delta^{b_2}_{a_1}\delta^{b_4}_{a_3}, \quad   \bar{I}^1=\delta^{{\bar{b}_1}}_{\bar{a}_2}\delta^{{\bar{b}_3}}_{\bar{a}_4}, \quad
\quad I^2=\delta^{b_4}_{a_1}\delta^{b_2}_{a_3}, \quad   \bar{I}^2=\delta^{{\bar{b}_1}}_{\bar{a}_4}\delta^{{\bar{b}_3}}_{\bar{a}_2}.
\end{equation}
with
\begin{equation}
\Omega_{ij}(x,\bar x)={\cal F}^{(-)}_i(x){\cal F}^{(-)}_j(\bar x)+ {1 - c_{--}^2 \over c_{+-}^2} {\cal F}^{(+)}_i(x){\cal F}^{(+)}_j(\bar x)
~.
\end{equation}

The functions that appear here are given by:
\begin{equation}
\begin{split}
&{\cal F}^{(-)}_1(x)= x^{-2h}(1-x)^{h_{\hat{\theta}}-2h}
\, _2F_1\left( {1\over p},{- 1 \over p};1- {N\over p},x\right) , \\
& {\cal F}^{(+)}_1(x)=x^{h_{\hat{\theta}}-2h}(1-x)^{h_{\hat{\theta}}-2h}
 \, _2F_1\left({N-1 \over  p}, {N+1 \over  p}; 1 - {N\over p},x\right), \\
&{\cal F}^{(-)}_2(x) = {1 \over k} x^{1-2h}(1-x)^{h_{\hat{\theta}}-2h}
\, _2F_1\left(1+ {1 \over  p},1 - {1\over p};2 - {N \over p},x\right) , \\
&{\cal F}^{(+)}_2(x)=-N x^{h_{\hat{\theta}}-2h} (1-x)^{h_{\hat{\theta}}-2h}
 \, _2F_1\left({N-1\over p},{N+1\over p};{N \over p}, x\right) ,\\
\end{split}
 \end{equation}
Here,
\begin{equation}
h={N^2-1\over 2N(N+k)}, \quad h_{\hat{\theta}}={N\over N+k},\quad p= N+k,
\end{equation}
and
\begin{equation}
\begin{split}
&c_{--} = N {\Gamma\left({N \over p}\right) \Gamma\left({-N \over p}\right) \over \Gamma\left({1 \over p}\right) \Gamma\left({-1 \over p}\right)} \\
&c_{+-} = -N {\Gamma\left(N \over p\right)^2 \over \Gamma\left({N+1 \over p}\right) \Gamma\left({N - 1 \over p}\right)}
\end{split}
\end{equation}

\subsection{The fourth and fifth correlators}
\label{subsec45bratch}
We now turn to the computation of the correlators
\begin{equation}
\begin{split}
G_{\phi_{-}^2\phi_{+} }(x) \equiv  \langle \phi_{-}^2(\infty) \phi_{+}(1) \overline{\phi}_{+}(x) \overline{\phi}_{-}^2(0) \rangle. \\
G_{ \phi_{+}^2\phi_{-}}(x) \equiv  \langle \phi_{+}^2(\infty) \phi_{-}(1) \overline{\phi}_{-}(x) \overline{\phi}_{+}^2(0) \rangle. \\
\end{split}
\end{equation}
Recall our identification of the double-trace operator with the field
\be
\begin{split}
&\phi_{+}^2 \leftrightarrow \sqrt{2} ({\cal A}_k)_{[a_1,a_2]}^{[\bar{b_1}, \bar{b_2}]} ({\cal \bar{A}}_1)^{[a_1,a_2]}_{[\bar{b_1}, \bar{b_2}]}\\
&\phi_{-}^2 \leftrightarrow \sqrt{2} ({\cal A}_1)_{[a_1,a_2]}^{[\bar{b_1}, \bar{b_2}]}  ({\cal \bar{A}}_{k*})^{[a_1,a_2]}_{[\bar{b_1}, \bar{b_2}]}
\end{split}
\ee
In this expression, the anti-symmetric primary operators are normalized
to have a two point function
\be
\left\langle {\cal A}_{[a_1,a_2]}^{[\bar{b_1}, \bar{b_2}]}(x),  {\cal A}^{[a_3,a_4]}_{[\bar{b_3}, \bar{b_4}]}(0) \right\rangle_{\widehat{su}(N)_{k}} = \delta^{[a_1, a_2]}_{[a_3,a_4]} \delta^{[\bar{b_1}, \bar{b_2}]}_{[\bar{b_3}, \bar{b_4}]} {1 \over |x|^{4 (N-2)(N+1) \over N (N+k)}},
\ee
where the number in the exponent of $x$ is twice the dimension of the anti-symmetric representation at level $k$: $\Delta_{{\cal A}_k} = {(N-2)(N+1) \over N (N+k)}$ Our normalization of the anti-symmetrized Kronecker-delta functions is as follows:
\begin{equation}
\begin{split}
 \delta^{[a_1, a_2]}_{[a_5, a_6]} &= {1 \over 2} \left(\delta^{a_1}_{a_5} \delta^{a_2}_{a_6} - \delta^{a_1}_{a_6} \delta^{a_2}_{a_5} \right)\\
\end{split}
\end{equation}
 Note that with the normalizations above, the fields $\phi_{-}^2$ and $\phi_{+}^2$ have two-point functions that are normalized to $2$ rather than $1$.

It is clear that the main building block we need is the following correlator in the $\widehat{su}(N)_1$ representation.
\begin{equation}
\label{sun1fourth}
{\cal H}^{a_1 a_2 a_4 \bar{b_3} \bar{b_5} \bar{b_6}}_{a_3 a_5 a_6 \bar{b_1} \bar{b_2} \bar{b_4}} = \langle {\cal A}^{[a_1,a_2]}_{[\bar{b_1}, \bar{b_2}]}(0), g_{a_3}^{\bar{b_3}}(x), {\bar{g}}^{a_4}_{\bar{b_4}}(1),  {\cal \bar{A}}_{[a_5,a_6]}^{[\bar{b_5}, \bar{b_6}]}(\infty) \rangle_{\widehat{su}(N)_1}
\end{equation}
where unbarred (barred) indices transform under left (right) affine transformations, raised indices transform as anti-fundamentals of $SU(N)$, and lowered indices transform as fundamentals of $SU(N)$. ${\cal A}$ is the primary field of the WZW model that transforms as an anti-symmetric tensor
on both the left and the right, while $g$, as above, transforms as a fundamental.

 In fact, since the level is $1$,  only one conformal block
contributes to this correlator. Consequently, we can write it as a product of a holomorphic and anti-holomorphic moving part ${\cal F}$ and
${\cal \bar{F}}$. The holomorphic part satisfies the Knizhnik-Zamolodchikov equation, which we write following \cite{DiFrancesco:1997nk} as
\begin{equation}
\label{kz}
\left[ \partial_x + {1 \over N + 1} \sum_a {t^{\alpha}_1 \otimes t^{\alpha}_2 \over x} + {1 \over N + 1} \sum_a {t_2^{\alpha} \otimes t_3^{\alpha} \over x - 1} \right] F^{a_4 a_1 a_2}_{a_3 a_5 a_6}(x) = 0.
\end{equation}
In this equation ${\alpha}$ is an index in the adjoint of $SU(N)$ and by $t^{\a ,}_i$ we mean the generator corresponding to $\a$
in the representation for the field $i$. This generator is, of course, a matrix.
In fact, we can guess the tensor structures that will occur in this equation and write
\begin{equation}
F^{a_4 a_1 a_2}_{a_3 a_5 a_6}(x) = F_1(x) \left(\delta_{a_3}^{a_1} \delta^{[a_2, a_4]}_{[a_5, a_6]} - \delta^{a_2}_{a_3} \delta^{[a_1, a_4]}_{[a_5, a_6]} \right) +  F_2(x) \delta^{a_4}_{a_3} \delta^{[a_1, a_2]}_{[a_5,a_6]}
 \equiv F_1(x) I_1 + F_2(x) I_2.
\end{equation}

We now need to evaluate the action of the group generators on the tensor structures $I_1$ and $I_2$. Note that the
generators in the fundamental representation are represented just by $t^{{\a ,} i}_j$ and so
\begin{equation}
(t_2^{\a})^i_j = t^{\a, i}_j, \quad (t_3^{\a})^i_j = t^{\a, i}_j.
\end{equation}
The anti-symmetric representation
has generators
\begin{equation}
(t_1^{\a})^{i,j}_{k,l} = {1 \over 2} \left(\delta^i_k t^{{\a ,} j}_l + \delta^j_l t^{{\a ,} i}_k - \delta^{{\a ,} i}_l t^j_k - \delta^j_k t^{{\a ,} i}_l\right)
\end{equation}
The main identity that we need is
\begin{equation}
\sum_{\a} t^{\a, i}_{k} t^{\a, j}_l = \delta^i_l \delta^j_k - {1 \over N} \delta^i_k \delta^j_l
\end{equation}

After some algebra, these identities lead to
\begin{equation}
\begin{split}
& \sum_{\a} (t^{\a}_1 \otimes t^{\a}_2) I_1 = {(N+1) (N-2) \over N} I_1\\
& \sum_{\a} (t^{\a}_1 \otimes t^{\a}_2) I_2 = -I_1 - {2 \over N} I_2 \\
& \sum_{\a} (t^{\a}_2 \otimes t^{\a}_3) I_1 = -{1 \over N} I_1 - 2 I_2 \\
& \sum_{\a} (t^{\a}_2 \otimes t^{\a}_3) I_2 = {N^2 - 1 \over N} I_2 \\
\end{split}
\end{equation}

Substituting this in the differential equation \eqref{kz}, we find the equations
\begin{equation}
\begin{split}
&\left(\frac{(N-2)(N+1)}{x}+\frac{1}{1- x}\right)
   F_1(x)+N (N+1) F_1'(x)=\frac{N F_2(x)}{x}, \\
&-2 N x F_1(x)+\left(\left(N^2-3\right) x+2\right)
   F_2(x)+N (N+1) (x-1) x F_2'(x)=0.
\end{split}
\end{equation}
Solving first for $F_1$, we find
\begin{equation}
\label{f1sol}
F_1(x)=  (x-1)^{\frac{1}{N}} x^{\frac{2}{N}-1}
   \left(c_1 + c_2 \frac{(N+1)
     x^{\frac{N-1}{N+1}}}{N-1} \,   _2F_1\left(\frac{N-1}{N+1},\frac{N+2}{N+1};\frac{2
   N}{N+1};x\right)\right)
\end{equation}
where $c_1$ and $c_2$ are coefficients. To determine them, we recognize that as we take $x \rightarrow 0$, the fields $g$ and ${\cal A}$ must fuse to give a primary that transforms in the anti-fundamental. This simplification occurs because, in our case, the level is $1$. The tensor structure corresponding to this fusion is $I_1$ and this tells us that the small $x$ behaviour of $F_1$ must be
\begin{equation}
F_1(x) \underset{x \rightarrow 0}{\longrightarrow} x^{-\Delta_{\cal A} \over 2} = x^{-{C_2(\text{\bf asym}) \over N + 1}} = x^{-1 + {2 \over N}}.
\end{equation}
This immediately tells us that $c_2 = 0$.

We can now write down a simple solution for $F_2$:
\begin{equation}
\label{f2sol}
F_2(x) = c_1 (x-1)^{\frac{1}{N}-1} x^{2/N}
\end{equation}
To fix $c_1$, we use the fact that as $x \rightarrow 1$, the two $g$ fields in \eqref{sun1fourth} must fuse to give the
identity. Once again, for general $k$, they could also have given the adjoint but we are interested in $k = 1$. This leads
to
\begin{equation}
F_2(x) \underset{x \rightarrow 0}{\longrightarrow} (1-x)^{-{2 C_2(\rep{f}) \over N + 1}} = (1 - x)^{N-1 \over N}
\end{equation}
which leads to
\begin{equation}
 c_1 = 1.
\end{equation}

Substituting this, we obtain the correlator \eqref{sun1fourth}
\begin{equation}
{\cal H}^{a_1 a_2 a_4 \bar{b_3} \bar{b_5} \bar{b_6}}_{a_3 a_5 a_6 \bar{b_1} \bar{b_2} \bar{b_4}} =  |1 - x|^{-2 + {2 \over N}} |x|^{-2 + {4 \over N}} \left( (-1 + x)  \left(\delta_{a_3}^{a_1} \delta^{[a_2, a_4]}_{[a_5, a_6]} - \delta^{a_2}_{a_3} \delta^{[a_1, a_4]}_{[a_5, a_6]} \right)  + x   \delta^{a_4}_{a_3} \delta^{[a_1, a_2]}_{[a_5,a_6]} \right)
\end{equation}

We are almost done. To get the coset correlator we need to contract this with two point functions that come from
the other elements of the coset. The coset correlator is given by
\begin{equation}
\begin{split}
G_{\phi_{+} \phi_{-}^2}(x) &= 2 {\cal H}^{a_1 a_2 a_4 \bar{b_3} \bar{b_5} \bar{b_6}}_{a_3 a_5 a_6 \bar{b_1} \bar{b_2} \bar{b_4}}   \times { \left\langle {\cal A}_{[a_1,a_2]}^{[\bar{b_1}, \bar{b_2}]}(0),  {\cal A}^{[a_5,a_6]}_{[\bar{b_5}, \bar{b_6}]}(\infty) \right\rangle_{\widehat{su}(N)_{-k-1-2N}} \over N^2(N-1)^2} \times  {\big\langle g^{a_3}_{\bar{b_3}}(x), g_{a_4}^{\bar{b_4}}(1) \big\rangle_{\widehat{su}(N)_k} \over N^2}\\
&=  {8 \over N^2} \left|(1-x)^{-\Delta_+} x^{\frac{2}{N}-1} \left(1 + {N - 2 \over 2} x\right)\right|^2,
\end{split}
\end{equation}
where apart from some index contractions, we have combined the power of $(1 - x)$ coming from the $\widehat{su}(N)_k$ two point
function with the power of $(1-x)$ from the $\widehat{su}(N)_1$ four point function, which gives us just what is required to match
the answer from the Coulomb gas formalism above.

The fifth correlator involves an almost identical computation in $\widehat{su}(N)_1$. We only need to change the two point
functions that we contract this result with, and this leads to
\begin{equation}
G_{\phi_{+}^2 \phi_{-}}(x) =  {8 \over N^2} \left|(1-x)^{-\Delta_-} x^{\frac{2}{N}-1} \left(1 + {N - 2 \over 2} x\right)\right|^2.
\end{equation}

%%% Local Variables:
%%% mode: latex
%%% TeX-master: "wncorrelators_paper"
%%% End:

\section{Group Theoretic Details \label{sec:grouptheory}}

\subsection{Orthogonal Basis}
In this appendix, we provide some useful group theoretic details. First, let us remind the reader that representations of $SU(N)$
are described by $N-1$ Dynkin labels. The Dynkin labels $d_i$ can be used to construct a Young tableaux for the $SU(N)$ 
representation as shown in Figure \ref{youngtableaux}
\begin{figure}[!h]
\label{youngtableaux}
\begin{center}
\includegraphics[width=0.6\textwidth]{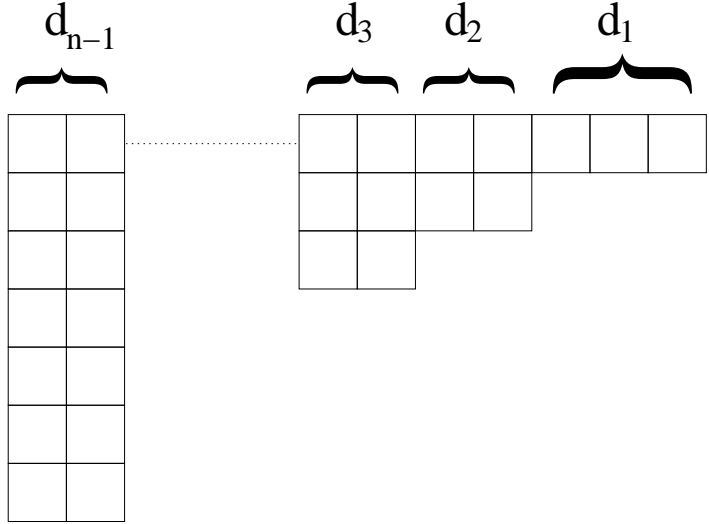}
\caption{Young Tableaux for SU(N)}
\end{center}
\end{figure}

We can instead also describe this Young tableaux by giving the number
of boxes in each row
\begin{equation}
r_i = \sum_{j=i}^N d_i, \quad i = 1 \ldots N-1
\end{equation}
The so-called orthogonal basis just counts the number of rows shifted by
an overall constant:
\begin{equation}
o_i = r_i - \sum_{i=1}^{N-1} r_i, \quad i = 1 \ldots N,
\end{equation}
with $r_N \equiv 0$. 

For some common representations, the orthogonal labels are
\begin{equation}
\begin{array}{ll}
\text{fundamental}&o_1 = {1 - {1 \over N}}, o_2 = o_3 = \ldots o_N = {-1 \over N} \\
\text{anti-fundamental}&o_1=o_2=\ldots o_{N-1} = 1 - {1 \over N}, o_{N-1} = {-1 \over N} \\
\text{adjoint} & o_1=1, o_2 = o_3 = \ldots o_{N-1} = 0, o_N = -1
\end{array}
\end{equation}
They Weyl vector has elements in the orthogonal basis given by
\begin{equation}
\text{Weyl~vector}:~o_i = {N + 1 \over 2} - i
\end{equation}

One of the many nice features of the orthogonal basis is the inner product
is orthogonal in this basis. For a representation $\rep{\Lambda}$ with 
orthogonal labels $o_i$, we have
\begin{equation}
\langle \rep{\Lambda}, \rep{\Lambda}\rangle = \sum_{i=1}^N o_i^2
\end{equation}
This fact is quite convenient for computing the Casimir of various representations, which is given by:
\be
C_2(\rep{\Lambda}) = {1 \over 2} \langle \rep{\Lambda}, \rep{\Lambda} + 2 \weyl{\rho} \rangle,
\ee
where $\weyl{\rho}$ is the Weyl vector.

Another nice feature of the orthogonal basis is that the Weyl group simply
acts by permuting the various elements. In particular, it is clear the order
of the Weyl group is $N!$. This is useful for computing characters. 

\subsection{Branching Function}
In this subsection, we provide some explicit formulas about the branching function for
representations of the coset ${\widehat{su}(N)_k \oplus \widehat{su}(N)_1 \over \widehat{su}(N)_{k+1}}.$ 

The branching function is given by
\begin{equation}\label{brfn}
b_{(\Lambda_+;\Lambda_-)}(q) =
 {1 \over \eta(q)^{N-1}} \sum_{\widehat{w} \in \widehat{W}} 
 \epsilon(\widehat{w}) q^{{1 \over 2 p(p+1)} ( (p+1)\widehat{w}(\Lambda_+ + \rho) - p (\Lambda_- + \rho) )^2}\ .
\end{equation}
Here $p=k+N$, and the sum is over the full affine Weyl group. The affine Weyl group is the 
semidirect product of the finite Weyl group and translations by elements ${\bf P}$ of the root lattice, 
and its action is given by
\begin{equation}
\widehat{w}(\rep{\Lambda} + \weyl{\rho}) = w (\rep{\Lambda}+\weyl{\rho}) + (k+N) {\bf P} \ ,
\end{equation}
where $w$ is an element of the ordinary $SU(N)$ Weyl group and $P$ is an 
element of the root lattice. In the orthogonal basis $P$ is quite simple to characterize: it is any vector with composed of $N$-integers that sum to zero.

With some manipulation, we can write
\begin{equation}
\label{brfnsimp1}
\begin{split}
b_{(\Lambda_+;\Lambda_-)}(q) &= q^{-c \over 24} \left[\prod_{s=1}^{N-1} \prod_{n=s}^{\infty} {1 \over 1 - q^n} \right] \\ &\times { \sum_{\widehat{w} \in \widehat{W}} \epsilon(\widehat{w}) q^{{1 \over 2 p(p+1)} \left[ ((p+1)\widehat{w}(\Lambda_+ + \rho) - p (\Lambda_- + \rho) )^2 - \rho^2 \right]} \over \sum_{w \in W} \epsilon(w) \, q^{{\langle \rho - w(\rho), \rho\rangle}}},
\end{split}
\end{equation}
where, in the denominator, we have the sum overe the ordinary Weyl group of $SU(N)$.
Hence, we see that the branching function automatically gives a regulated version of $Z_{\rm hs}$ in \eqref{eq:bulkpartfn}. What is of
interest is the part of the branching function, which is on the second line. Ideally we would like that when we sum over all representations as in \eqref{eq:zcftschem} we get exactly the $Z_{\rm scal}$ term in \eqref{eq:bulkpartfn}. However, while we do get this term
as we have explained in the text, we also get additional terms.

%%% Local Variables: 
%%% mode: latex
%%% TeX-master: "wncorrelators_paper"
%%% End: 

\bibliographystyle{JHEP}
\bibliography{references}
\end{document}